\documentclass{aa}  

\usepackage{graphicx}
\usepackage{txfonts}
\usepackage{hyperref}

\usepackage{xcolor} 
\usepackage{txfonts} 
\usepackage{rotating} 
\usepackage{url}  
\usepackage{mathtools, amsmath, amssymb, amsfonts} 
\usepackage{ccicons} 
\usepackage{mathabx}
\usepackage{multirow}

\usepackage{subcaption}

\DeclareUnicodeCharacter{2212}{-}

\graphicspath{{Figures/}}

\defcitealias{2021ApJ...910...51K}{Kom21}
\defcitealias{2021MNRAS.508.3611P}{Pic21}

\begin{document}

   \title{Disc population synthesis: Decrease in the solid mass reservoir through pebble drift}
   
   \titlerunning{Disc population synthesis}

   \author{J. Appelgren
          \inst{1}
          ,
          M. Lambrechts\inst{2,1}
          \and
          N. van der Marel\inst{3}
          }

   \institute{
    \inst{1} Lund Observatory, Department of Astronomy and Theoretical Physics, Lund University, Box 43, 22100 Lund, Sweden \\
    email: johan@astro.lu.se \\
    \inst{2} Center for Star and Planet Formation, GLOBE Institute, University of Copenhagen, Øster Voldgade 5-7, 1350 Copenhagen, Denmark \\
    \inst{3} Leiden Observatory, Leiden University, P.O. Box 9531, NL-2300 RA Leiden, the Netherlands \\         
             }

   \date{Received ; accepted }

\abstract{Surveys of star-forming regions reveal that the dust mass of protoplanetary discs decreases by several orders of magnitude on timescales of a few million years. This decrease in the mass budget of solids is likely due to the  radial drift of millimetre (mm) sized solids, called pebbles, induced by gas drag. However, quantifying the evolution of this dust component in young stellar clusters is difficult due to the inherent large spread in stellar masses and formation times. Therefore, we aim to model the collective evolution of a cluster to investigate the effectiveness of radial drift in clearing the discs of mm-sized particles. We use a protoplanetary disc model that provides a numerical solution for the disc formation, as well as the viscous evolution and photoevaporative clearing of the gas component, while also including the drift of particles limited in size by fragmentation. We find that discs are born with dust masses between 50\,M$_\Earth$ and 1000\,M$_\Earth$, for stars with masses, respectively, between 0.1\,M$_\odot$ and 1\,M$_\odot$. The majority of this initial dust reservoir is typically lost through drift before photoevaporation opens a gap in the gas disc for models both with and without strong X-ray-driven mass-loss rates. We conclude that the decrease in time of the mass locked in fragmentation-limited pebbles is consistent with the evolution of dust masses and ages inferred from nearby star-forming regions, when assuming viscous evolution rates corresponding to mean gas disc lifetimes between 3\,Myr and 8\,Myr.

}

 


   \maketitle
%

\section{Introduction}

Surveys of young star-forming regions, such as the 1\,Myr-old Perseus cluster, reveal typical dust disc masses that exceed $100$\,M$_\Earth$ \citep{2020A&A...640A..19T, 2020ApJ...890..130T}. Older star-forming regions, on the other hand, often show vastly reduced dust masses by up to two orders of magnitude, which are albeit characterised by a large spread \citep{2013ApJ...771..129A, 2016ApJ...828...46A, 2016ApJ...831..125P,  2018ApJ...860...77E, 2018MNRAS.478.3674R,2018ApJ...859...21A, 2019MNRAS.482..698C, 2019A&A...626A..11C, 2020ApJ...890..130T, 2021ApJ...913..123G}. However, care should be taken in the interpretation of individual dust disc mass measurements, as these depend on model choices such as the chosen opacity prescription, particle size distribution, and assumptions of optically thin continuum emission \citep{2019ApJ...877L..22L, 2019ApJ...877L..18Z}.

The dominant process responsible for removing dust is not yet completely clear. Two likely candidates are the radial drift of dust pebbles and the formation of planetesimals and planets. Radial drift occurs due to gas drag from gas on sub-Keplerian orbits removing angular momentum from the pebbles, which causes them to drift inwards in the disc \citep{1977MNRAS.180...57W}. If dust particles are sufficiently large, this occurs in most regions of protoplanetary discs but can be halted at locations with a pressure bump where gas drag is not present \citep{2008A&A...480..859B,  2016ApJ...831..125P}. In \cite{2020A&A...638A.156A}, we studied the effect of radial drift in population synthesis models and found radial drift to be compatible with typical measurements of disc dust masses. However, this study only traced a single dust size and did not include gas disc clearing by photoevaporation. The effect of limiting dust growth by fragmentation and the impact of photoevaporation is the focus of the present population synthesis study. 

Estimates of dust grain sizes in protoplanetary discs have found maximum sizes on the order of mm-cm, based on measured opacity slopes \citep{2012ApJ...760L..17P, 2015ApJ...813...41P, 2016A&A...588A..53T, 2019ApJ...883...71C}. However, if maximum dust sizes are estimated from the degree of polarisation in the emission from protoplanetary discs, typically lower maximum sizes of 0.10-0.15\,mm are found \citep{2015ApJ...809...78K, 2016ApJ...831L..12K}, although 1\,mm is still possible for very low turbulence values \citep{2021ApJ...913..117U}. Dust coagulation is a complex process whereby the collision of two particles can result in very different outcomes, including sticking, bouncing, fragmentation, mass transfer, cratering, and erosion, depending on the sizes and velocities of the two particles \citep{2008ARA&A..46...21B, 2010A&A...513A..57Z, 2011ApJ...731...95O, 2011ApJ...731...96O, 2018SSRv..214...52B}. The outcome of these coagulation processes sets the size, and therefore Stokes number, of the largest particles that characteristically hold a very large fraction of the total disc dust mass \citep{2010A&A...513A..79B}. Importantly, the dominant particle size affects not only the speed of the radial drift of these particles but also their ability to form planetesimals via the streaming instability \citep{2005ApJ...620..459Y, 2007Natur.448.1022J, 2015A&A...579A..43C, 2017A&A...606A..80Y, 2021MNRAS.503.5254L} and the efficiency of pebble accretion \citep{2010A&A...520A..43O, 2012A&A...544A..32L, 2014A&A...572A.107L}.

Protoplanetary discs are formed together with their host star when dense cores of giant molecular clouds collapse under their own gravity. The formation of the disc is thought to last a few hundred thousand years up to $\sim1$\,Myr \citep{2005A&A...442..703H, 2007ARA&A..45..565M}. As the disc continues to evolve, gas accretion removes gas from the disc over a timescale of a few to several million years \citep{1998ApJ...495..385H, 2022arXiv220309930M}. The accretion of gas is driven by a redistribution of angular momentum in the discs. The mechanism behind the redistribution is not yet fully understood, but it is believed to be disc turbulence driven by the magnetorotational instability (MRI)\citep{1998RvMP...70....1B} or disc winds \citep{2013ApJ...769...76B}, or a combination of the two. For a recent review on magneto-hydrodynamic disc processes, we refer to \citet{2022arXiv220309821L}. Due to the poor understanding of what drives the turbulence, this is commonly parameterised by the well-known $\alpha$-disc model \citep{1973A&A....24..337S}. During this time, gas accretion rates drop from as high as $10^{-6} \ \mathrm{M}_\odot$/yr, for the youngest discs, to $\lesssim 10^{-9} \ \mathrm{M}_\odot$/yr for discs as old as a few million years \citep{2016A&A...591L...3M}. Simultaneously, the gas disc mass is reduced from $\sim 0.1\ \mathrm{M}_\odot$ to $\sim 0.001\ \mathrm{M}_\odot$ \citep{2022arXiv220309818M}. However, estimating gas disc masses from, for example, CO gas emission, is highly uncertain. Therefore, dust disc masses have been used to infer gas disc masses, under the crude assumption of a global dust-to-gas ratio of 0.01 \citep[e.g.][]{2022arXiv220309930M}. This value is typical of the interstellar medium, but no longer valid when considering the effects of pebble drift \citep{2020A&A...638A.156A}. Furthermore, dust disc masses themselves may be underestimated if the millimeter emission is optically thick.

It is important to include the correct range of stellar masses when trying to model the evolution of a cluster of discs. Lower mass stars are more commonly formed than solar-mass stars according to the stellar initial mass function \citep{1955ApJ...121..161S, 2001MNRAS.322..231K, 2003PASP..115..763C}. The dominating presence of sub-solar stars is of importance because the mass of dust discs has been found to scale with the stellar mass in nearby star-forming regions as $M_\mathrm{dust} \propto M_\star^{1.3-2.4}$, although with a large scatter \citep{2016ApJ...831..125P, 2017AJ....153..240A}. Similarly, a trend has been found where higher stellar gas accretion rates relate to more massive dust discs, but again with a significant scatter \citep{2016A&A...591L...3M}.

The final dispersal of the gas component of protoplanetary discs is suggested to be driven by photoevaporative winds \citep[e.g.][]{1994ApJ...428..654H, 1994ApJ...429..781S, 2001MNRAS.328..485C}. A combination of extreme ultraviolet (EUV), X-ray, and far
ultraviolet  (FUV) radiation heats the gas at the disc surface to the point where the gas becomes gravitationally unbound and is lost from the disc. When the gas removal rate via photoevaporation and the gas accretion rate through the disc become comparable, the inner disc is no longer efficiently replenished by gas. This results in the efficient dispersal of the inner disc, creating an inner disc cavity that expands outwards with time.

Several prescriptions of photoevaporation aimed for use in population synthesis models have been developed. \cite{2012MNRAS.422.1880O} have provided a prescription for X-ray-driven photoevaporation, which was extended by an updated X-ray photoevaporation prescription \citep{2019MNRAS.487..691P, 2021MNRAS.508.1675E, 2021MNRAS.508.3611P}. \cite{2021ApJ...910...51K} have provided an alternative prescription for photoevaporation driven by  EUV, FUV, and X-rays, using a method similar to that of \cite{2018ApJ...865...75N}. The importance of each of these drivers has been the subject of a number of studies. In particular, EUV and X-rays are more straightforward to model, whereas FUV is a bigger challenge because it is affected by dust evolution and chemistry \citep{2015ApJ...804...29G}. \citet{2012MNRAS.422.1880O} concluded that X-rays are the dominant driver of photoevaporation. \cite{2019MNRAS.487..691P} similarly argued that X-rays dominate over EUV. In contrast, \cite{2018ApJ...865...75N} found that FUV-driven photoevaporation is dominant. We explore the impact of both the \cite{2021MNRAS.508.3611P} and the \cite{2021ApJ...910...51K} models in the following.

In this paper, we first elaborate on the method used for the population synthesis model (Sect. \ref{sec:model}). We then present the results of the study. We focus on the evolution of the disc dust mass and how its evolution is affected by the strength of the assumed disc viscosity and the two different photoevaporation prescriptions available in the literature (Sect. \ref{sec:results}). This is followed by a discussion of our results, their stellar mass dependency, and the implications on the gas and dust radius evolution with time (Sect. \ref{sec:discussion}). Finally, we present our conclusions (Sect. \ref{sec:conclusions}).

\section{Model} \label{sec:model}

In this paper, we present an update to the disc model used in \cite{2020A&A...638A.156A}. As in the previous paper, we calculate the evolution of discs around stars in a newly formed stellar cluster using viscous gas evolution and radial drift of solids. We briefly review the used methodology here. In addition, here we  evolve the size of the dust particles, while including the clearing of the gas disc by photoevaporation and  using a more complete temperature treatment.

\subsection{Disc formation}

The formation of the protoplanetary disc is modelled from the collapse of an over-dense Bonnor-Ebert sphere, similar to the work in \cite{2005A&A...442..703H, 2013ApJ...770...71T}. The collapse occurs inside-out with the collapse front expanding outwards over time. The infall rate of gas mass onto the disc is given by: 
\begin{align}
        \dot{M}_\mathrm{g, inf} = 4 \pi r_\mathrm{cf}( t )^2 \rho ( r_\mathrm{cf}( t ) ) \dfrac{\mathrm{d} r_\mathrm{cf}}{\mathrm{d} t}, \label{eq:mdotinf}
\end{align}
where $r_\mathrm{cf}$ is the radius of the collapse front, $\rho(r_\mathrm{cf})$ is the density of the cloud core at this radius and $\mathrm{d}r_\mathrm{cf}/\mathrm{d}t$ is the velocity at which it expands outwards. A more detailed description of how these quantities are calculated can be found in Appendix A of \cite{2020A&A...638A.156A}.

The gas is distributed across the disc according to the following equation: 
\begin{align}
        \dot{\Sigma}_\mathrm{g, inf}(r) = \dfrac{\dot{M}_\mathrm{g, inf}}{8\pi R_\mathrm{c}^2}\left( \dfrac{r}{R_c} \right)^{-3/2} \left[ 1 - \left(\dfrac{r}{R_\mathrm{c}}\right)^{1/2} \right]^{-1/2}. \label{eq:sigdotinf}
\end{align}
The infalling material lands on the disc within the centrifugal radius $R_\mathrm{c}$,  given by:
\begin{align}
        R_\mathrm{c} = \dfrac{\Omega_0^2 r_\mathrm{cf}\left( t \right)^4}{G M\left( r_\mathrm{cf}, \right)}, \label{eq:Rc}
\end{align}
where $\Omega_0$ is the rotation rate of the cloud core and $M(r_\mathrm{c})$ is the mass interior to the collapse front radius. We assume here that the angular momentum of the infalling material is conserved. This is a simplification that ignores the role of magnetic braking \citep[e.g.][]{2006ApJ...647..374G, 2019MNRAS.489.1719W} and possible anistropic accretion \citep[e.g.][]{2011MNRAS.411.1354S, 2017ApJ...846....7K}. However, a detailed modelling of the angular momentum of molecular cloud cores and their evolution during the collapse phase is beyond the scope of this paper.

\subsection{Gas disc evolution} \label{sec:mod:gas}

We evolved the disc viscously using the well-established alpha-disc prescription of \cite{1973A&A....24..337S}, where the viscosity of the disc is described by $\nu = \alpha c_\mathrm{s}^2/\Omega$. Here, $c_\mathrm{s}$ is the sound speed and $\Omega$ is the rotation rate in the disc, while $\alpha$ sets the strength of the viscosity. We treat $\alpha$ as the value which evolves the disc on a timescale consistent with measured disc lifetimes. We do not treat it as  a measurement of the true level of midplane turbulence stirring the dust midplane of the disc, which has been observationally inferred to be much lower in the disc \citep{2016ApJ...816...25P, 2020A&A...642A.164V, 2022ApJ...930...11V}. Instead, we parameterise the latter as $\alpha_\mathrm{t}$, which we also use to calculate the size of fragmentation-limited dust (see Sect. \ref{sec:dustsize}).

Using accretion rates of pre-main sequence stars, \cite{1998ApJ...495..385H} found  $\alpha = 10^{-2}$. However, recent estimates of line-width broadening by turbulent motions in outer disc regions have found values of $\alpha \lesssim 10^{-3}$ \citep{2018ApJ...864..133T, 2018ApJ...856..117F, 2020ApJ...895..109F}. Moreover, measurements of the fraction of stars with discs in star-forming regions suggest that the discs dissipate with a characteristic timescale of 2.5-3\,Myr \citep{2001ApJ...553L.153H, 2009AIPC.1158....3M}. Such disc lifetimes of $\sim 3$\,Myr would suggest high $\alpha$-values to drive  disc evolution. However, more recent measurements of the disc fraction finds that, in those star-forming regions that are not subject to external photoevaporation, the characteristic timescale for disc dissipation is $\sim 8$\,Myr \citep{2021ApJ...921...72M, 2022ApJ...939L..10P}. The slower disc evolution (indicated by the longer disc dissipation timescale) would suggest lower values of $\alpha$. We used a base value, which we labelled as $\alpha_\nu$, of either $10^{-2}$ or $10^{-3}$  to explore both scenarios of faster and slower disc evolution. In practise, $\alpha$ is calculated as: 
\begin{align}
    \alpha = \alpha_\nu + e^{-Q^4}, \label{eq:alpha_nu}
\end{align}
which increases the disc viscosity when the disc becomes
gravitationally unstable following \citet{2010ApJ...713.1134Z}. Here, Q is the Toomre parameter expressing the gravitational stability of the disc.

The viscous evolution of the disc is calculated using the equation from \cite{1981ARA&A..19..137P}. The total surface density evolution of the disc is then given by: 
\begin{align}
        \dfrac{\partial\Sigma_\mathrm{g}}{\partial r} = \dfrac{3}{r} \dfrac{\partial}{\partial r} \left[ r^{1/2} \dfrac{\partial}{\partial r} \left( \nu \Sigma_\mathrm{g} r^{1/2} \right)  \right] +\dot{\Sigma}_\mathrm{g, inf}\left(r\right) - \dot{\Sigma}_\mathrm{g, PE,}\left(r\right), \label{eq:SigEvo}
\end{align}
where $\dot{\Sigma}_\mathrm{g, PE}$ is the mass loss due to internal photoevaporation. The way we calculate this is explained in Sect. \ref{subsec:model:pe} and Appendix\,\ref{app:model:PE}.

\subsection{Dust evolution}

The infall rate of dust onto the disc is the infall rate of gas multiplied by the assumed dust-to-gas ratio of the cloud core. We assume this to be uniformly $Z=0.01$. 

\subsubsection{Dust drift}

The evolution of the dust in the disc is governed by the radial gas flow and  the radial drift of the dust particles. Radial drift occurs because the dust feels a slight headwind from the gas that orbits at a sub-Keplerian velocity \citep{1977MNRAS.180...57W}. This headwind removes angular momentum from the dust causing it to drift inwards in the disc. The speed at which the dust drifts is given by:
\begin{align}
    v_\mathrm{drift} = - \dfrac{2\tau_\mathrm{s}}{1 + \tau_\mathrm{s}^2} \eta v_\mathrm{k}. \label{eq:dustdrift}
\end{align}
Here, $v_\mathrm{k}$ is the Keplerian orbital velocity and $\eta$ is a measure of the radial pressure support in the disc, given by:
\begin{align}
    \eta = -\dfrac{1}{2}\left(\dfrac{H}{r}\right)^2 \dfrac{\mathrm{d} \ln P}{\mathrm{d} \ln r}. \label{eq:eta}
\end{align}
The Stokes number, $\mathrm{St}$, of the dust particles is a measure of how aerodynamically coupled particles are to the gas. The ratio $H/r$ expresses the gas disc aspect ratio. To calculate the Stokes number we take into account the Epstein and the Stokes regimes, given by:
\begin{align}
    \mathrm{St} =  \begin{cases}
    \dfrac{\sqrt{2\pi}\rho_\bullet a_\mathrm{d}}{\Sigma_\mathrm{g}} \ &\mathrm{if} \ a_\mathrm{d} < \dfrac{9}{4} \lambda_\mathrm{mfp}, \\
    \dfrac{4}{9} \dfrac{a_\mathrm{d}}{\lambda_\mathrm{mfp}} \dfrac{\sqrt{2\pi}\rho_\bullet a_\mathrm{d}}{\Sigma_\mathrm{g}} \ &\mathrm{if} \ a_\mathrm{d} \geq \dfrac{9}{4} \lambda_\mathrm{mfp}. \label{eq:stokes}
\end{cases}
\end{align}
Here, $\rho_\bullet$ is the material density of the dust particle which we set to 1.6 g/cm$^3$ \citep{2012A&A...539A.148B}. The size of the dust grains, $a_\mathrm{p}$, is discussed further in Sect. \ref{sec:dustsize}.

Because the dust is coupled to the gas, small grains simply move along with the gas in the disc. Taking this into account, the total radial velocity of the dust, $v_\mathrm{d}$, is given by: 
\begin{align}
    v_\mathrm{d,r} = v_\mathrm{drift} + \dfrac{v_\mathrm{g}}{1 + \mathrm{St}^2}, \label{eq:vrdust}
\end{align}
where $v_\mathrm{g}$ is the radial velocity of the gas. The surface density of dust is then evolved with the following continuity equation:
\begin{align}
    \dfrac{\partial\Sigma_\mathrm{d}}{\partial t} = -\dfrac{1}{r}\dfrac{\partial}{\partial r}\left(r\Sigma_\mathrm{d} v_\mathrm{d,r}\right) + \Sigma_\mathrm{d,inf}.
\end{align}

\subsubsection{Dust sizes} \label{sec:dustsize}

We followed the evolution of the largest grains that carry most of the mass in typical dust size distributions \citep{2010A&A...513A..79B}. The dust grows from an initial monomer size of 1 $\mu m$ to the fragmentation limit on a timescale given by: 
\begin{align}
    t_\mathrm{grow} = \dfrac{\Sigma_\mathrm{g}}{\Sigma_\mathrm{d}\Omega_\mathrm{K}}. \label{eq:tgrow}
\end{align}
For grains that reach their maximum size limit by fragmenting, the maximum grain size is set via 
\begin{align}
a_\mathrm{frag} = f_\mathrm{f}\dfrac{2}{3\pi}\dfrac{\Sigma_\mathrm{g}}{\rho_\bullet\alpha_\mathrm{t}} \dfrac{u_\mathrm{f}^2}{c_\mathrm{s}^2} \label{eq:a_frag},
\end{align}
where $f_\mathrm{f} = 0.37$ is a fudge factor to better match detailed coagulation simulations \citep{2012A&A...539A.148B}. We set the level of the midplane turbulence $\alpha_\mathrm{t}$ to $10^{-4}$, in line with the lower $\alpha$ values inferred from observations \cite{2016ApJ...816...25P, 2022ApJ...930...11V}. We note that we treat it as a separate parameter from $\alpha_\nu$ (Eq.\,\ref{eq:alpha_nu}). The fragmentation threshold velocity, $u_\mathrm{f}$, depends on the material properties of the grains. Laboratory experiments of silicate dust grain collisions suggest limits of about 1\,m/s \citep{2008ARA&A..46...21B}.  Water ice was thought to be stickier than silicates and to have a fragmentation speed of about 10\,m/s \citep{2015ApJ...798...34G}. However, more recent studies have concluded that the fragmentation speed of water ice is likely to be lower. The stickiness of CO$_2$-ice is also thought to be quite low, with fragmentation speeds of about 1\,m/s \citep{2016ApJ...818...16M}. Similar results are found for mixtures of water and CO$_2$ ices \citep{2016ApJ...827...63M}. Therefore, the choice of 1\,m/s as a global fragmentation velocity should be suitable.

Particles that are not limited by fragmentation in their growth naturally approach their drift-limited sizes in our model. Their maximum size is then bound by: 
\begin{align}
    a_\mathrm{drift} = f_\mathrm{d} \dfrac{2\Sigma_\mathrm{d}}{\pi\rho_\bullet}\dfrac{v_\mathrm{k}^2}{c_\mathrm{s}^2}\gamma^{-1},
\end{align}
where $f_\mathrm{d} = 0.55$ is a fudge factor with similar purpose as $f_\mathrm{f}$ and $\gamma = |\frac{\mathrm{d} \ln P}{\mathrm{d} \ln r}|$ \citep{2012A&A...539A.148B}.

\subsection{Disc temperature}

 The temperature of the disc was calculated similarly to the work in \citet{2016MNRAS.461.2257K}. We included irradiation and viscous heating. The temperature due to irradiation is given by  \cite[e.g.][]{2004ApJ...606..520M}:
 \begin{align}
        \sigma_\mathrm{SB} T_\mathrm{irr, \star}^4 = \dfrac{L_\star}{4\pi r^2}\dfrac{H}{r}\left(\dfrac{\mathrm{d} \ln H}{\mathrm{d} \ln r} - 1 \right), \label{eq:IrrFluxStar}
\end{align}
where $L_\star$ is the stellar luminosity. We assume that the luminosity of the star scales linearly as $L_\star = L_\odot(M_\star/\mathrm{M}_\odot)$ \citep{2002A&A...382..563B, 2019A&A...632A...7L}. 

The heating rate due to irradiation is given by \cite[e.g.][]{2004ApJ...606..520M}:
\begin{align}
        \Gamma_\mathrm{irr} = \dfrac{8\sigma_\mathrm{SB} \left( T_\mathrm{irr, \star}^4 + T_\mathrm{core}^4 \right)}{ \tau/2 + 1/\sqrt{3} + 1/(3\tau)}, \label{eq:IrrHeatRate}
\end{align}
where $T_\mathrm{core} = 10$ K is the temperature of the cloud core, $\tau = \kappa \Sigma/2$ is the optical depth, and $\kappa$ is the opacity.

The heating rate due to viscous heating is given by \cite[e.g.][]{2016MNRAS.461.2257K}:
\begin{align}
        \Gamma_\mathrm{Visc} = \dfrac{9}{4}\nu \Sigma_\mathrm{g}\left(r \dfrac{\partial\Omega}{\partial r}\right)^2. \label{eq:ViscHeatRate}
\end{align}
Heat is lost from the disc by radiative cooling  \cite[e.g.][]{2004ApJ...606..520M}:
\begin{align}
        \Lambda_\mathrm{rad} = \dfrac{8\sigma_\mathrm{SB}T_\mathrm{mid}^4 }{3\left( \tau/2 + 1/\sqrt{3} + 1/(3\tau) \right)}, \label{eq:cooling}
\end{align}
where $T_\mathrm{mid}$ is the midplane temperature of the disc. The midplane temperature due to irradiation and viscous heating is given by the balance of the heating and cooling rates. 

We used an opacity prescription similar to that in \cite{2016MNRAS.461.2257K}, where the opacity is that of solid grains and given by $\kappa = 0.052\ \mathrm{cm}^2/g \left(T/K\right)^{0.738}$ \citep{2009ApJ...694.1045Z}. This opacity scaling assumed a solar composition of dust grains. However, above a temperature of $\sim 1500$ K, most dust will have sublimated. This will cause the opacity to drop and the disc to cool until the dust solidifies again. Unless the much lower gas opacity is enough to drive the disc to higher temperatures, this results in a temperature plateau around 1500 K. This effect has been found in more complex disc temperature models \citep[see e.g.][]{2010A&A...513A..79B, 2019ApJ...881...56S, 2021MNRAS.503.5254L}. We therefore imposed a temperature limit of 1500 K in our model.

\subsection{Photoevaporation} \label{subsec:model:pe}

\begin{figure}[ht]
    \centering
    \includegraphics[width=\hsize]{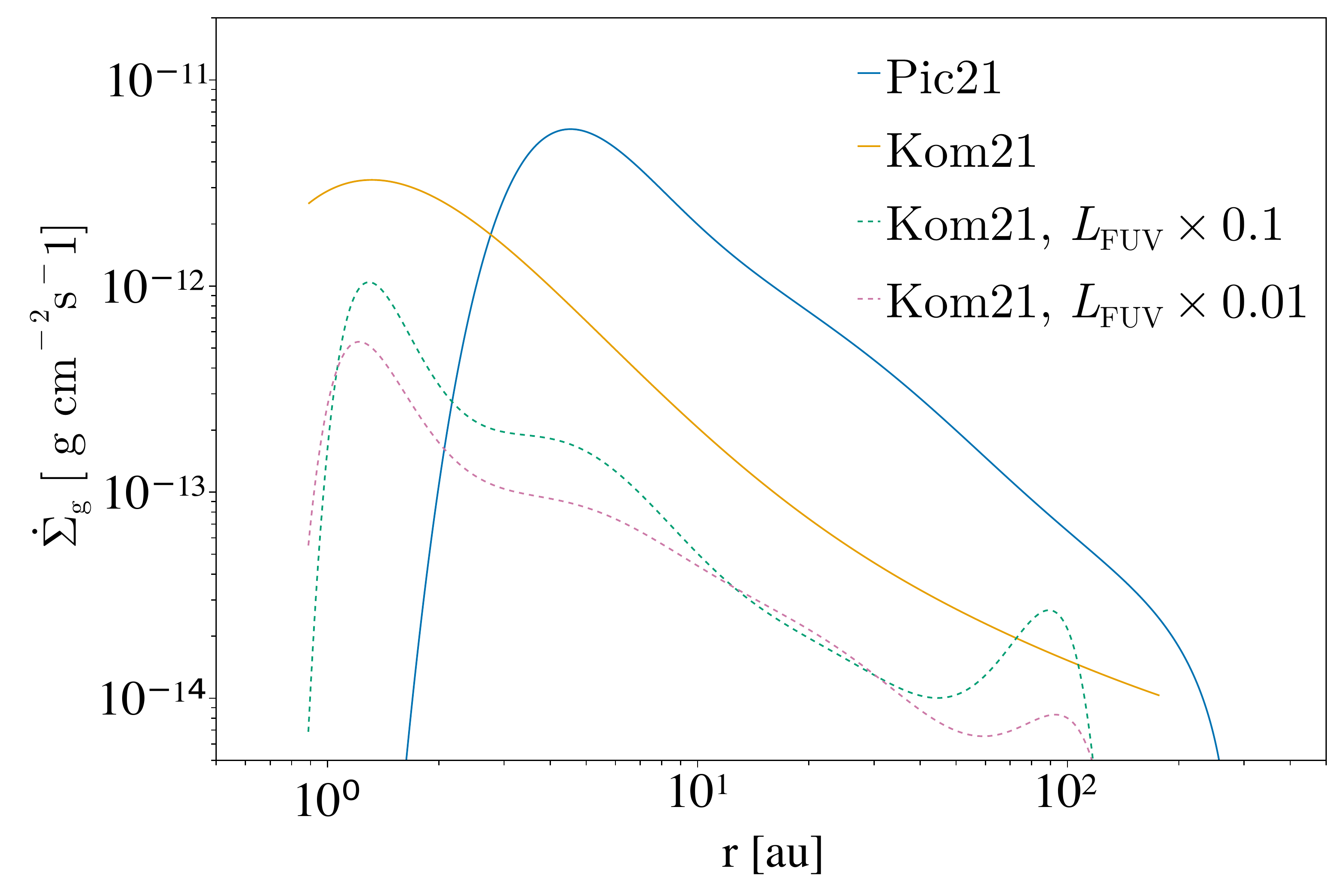}
    \caption{Photoevaporation rates for a $1\ \mathrm{M}_\odot$-star using identical X-ray luminosity for both the \citetalias{2021MNRAS.508.3611P} (blue) and \citetalias{2021ApJ...910...51K} (yellow) prescriptions. The two dashed lines show the \citetalias{2021ApJ...910...51K} models where the FUV luminosity has been reduced by a factor of 10 and of 100.}
    \label{fig:meth:PE_sdot_diff}
\end{figure}

\begin{figure*}[ht]
\centering
\includegraphics[width=\hsize]{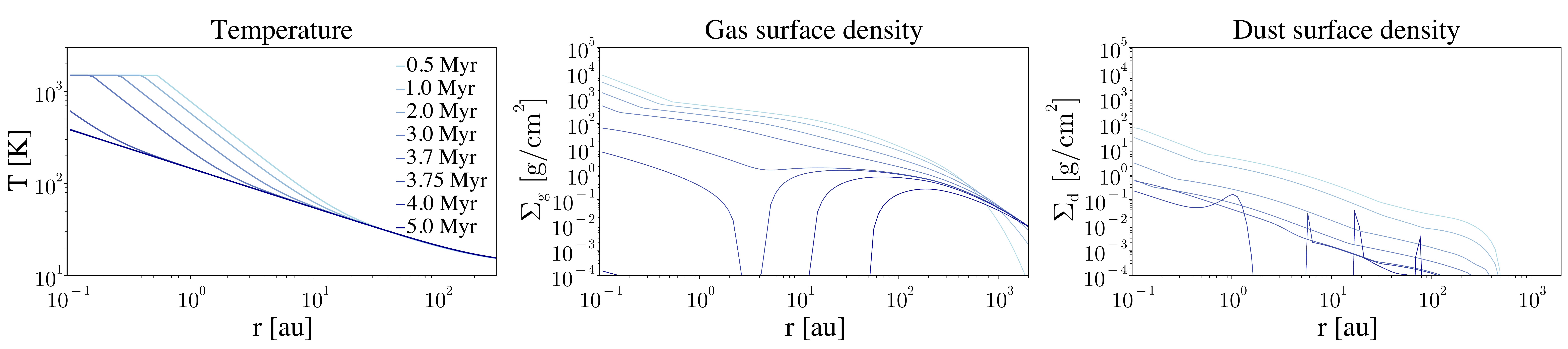}
      \caption{Temperature (left), gas surface density (middle), and dust surface density (right) evolution for a disc forming from a 1\,$\mathrm{M}_\odot$ cloud core with a centrifugal radius of 10\,au and photoevaporation according to \citetalias{2021MNRAS.508.3611P}. The dust is allowed to grow to the fragmentation limit. In the innermost part of the disc the temperature can reach 1500 K due to viscous heating and in the outer disc the temperature is set purely by irradiation. Photoevaporation begins to open up a gap in the gas disc around 3.75\,Myr. By 4\,Myr, the inner gas disc is completely dispersed, leaving behind an outer depleted disc. We find that 0.6 $M_\mathrm{Earth}$ of the dust is retained outside this gap and is pushed outwards with time.}
         \label{Fig:res:PPD_plots_nom}
\end{figure*}

Photoevaporation removes gas from the surface of the protoplanetary disc by either EUV, FUV, or X-rays heating the gas until it can escape the disc \citep{2009ApJ...705.1237G, 2015ApJ...804...29G, 2010MNRAS.401.1415O, 2011MNRAS.412...13O, 2012MNRAS.422.1880O, 2017ApJ...847...11W, 2018ApJ...857...57N, 2018ApJ...865...75N}. In this work, we look at two different photoevaporation prescriptions: one by \citet[][]{2021MNRAS.508.3611P} (hereafter Pic21) and one by \citet[][]{2021ApJ...910...51K} (hereafter Kom21).

The two prescriptions differ in the physics that is included, and also in the methods used to derive the photoevaporation rates. To model the photoevaporation, \citetalias{2021MNRAS.508.3611P} included X-rays and EUV radiation, although the X-ray component is the dominant contributor to the heating and mass loss rate \citep{2019MNRAS.487..691P}. The model from \citetalias{2021ApJ...910...51K} includes X-rays, EUV, and FUV radiation, with the FUV radiation being the dominant source of heating. In this model, the X-ray contribution is negligible (as illustrated in Fig.\,\ref{fig:meth:PE_sdot_diff}), where mass loss rates are driven by the FUV radiation, even if the FUV fraction would be reduced by a factor of 100.
We therefore chose to present both of these models to reflect the uncertainty in modeling photoevaporative mass loss (for a further discussion, see Sect.\,\ref{sec:diss:photovap}).

The equations and the parameters used to calculate the photoevaporation rates are given in Appendix \ref{app:model:PE}. To find the parameters for stellar masses that fall in between the ones given in either model, we linearly interpolated over the grid of masses. For stellar masses that fall outside the range of masses modelled in the prescriptions, we used the parameters for the lowest and highest mass given to avoid extrapolating. We did this because, in general, the parameters of either prescription are not monotonically increasing or decreasing with stellar mass. For the \citetalias{2021MNRAS.508.3611P} model, this upper mass is 1\,M$_\odot$ and stars above this mass will prove very uncommon in our sample; therefore, the choice of using the 1\,M$_\odot$ cut off will be of little consequence. With the  \citetalias{2021ApJ...910...51K} model the lower mass is 0.3\,M$_\odot$ and thus more stars will be affected since our stellar masses extend to $\lesssim 0.1 \ \mathrm{M}_\odot$, possibly leading to moderately overestimating the photoevaporation rate for very low-mass stars.

\subsection{Population synthesis} \label{sec:Model:popsynth}

The main cloud core parameters that are varied in the population synthesis are the mass and angular momentum. The masses of cores in molecular clouds are described by the so-called core-mass-function (CMF). This is similar in form to the initial mass function of stars (IMF) \citep{2007ARA&A..45..565M}. Therefore, we sampled masses of the cloud cores from the Kroupa IMF \citep{2001MNRAS.322..231K}. We limited our sampling to masses between 0.1--2\,$\mathrm{M}_\odot$.

To set the angular momentum of the cloud core, we set the maximum centrifugal radius, which is connected to the angular momentum of the cloud core through Eq. \ref{eq:Rc}. This is the largest radius at which any material will land on the disc during its formation. We select centrifugal radii from a normal distribution with a standard deviation of 30\,au and a mean scaled linearly with the cloud core mass as: 
\begin{align}
R_\mathrm{c} = 10 \ \mathrm{au} \ \left(M_\mathrm{core}/\mathrm{M}_\odot \right),
\end{align}
as in \citet[][specifically, their Appendix D]{2020A&A...638A.156A}. In order to limit the number of parameters varied in the population synthesis, we chose to use a fixed value for the temperature of the cloud core of $T_\mathrm{core} = 10$\,K across all simulations.

The stars in a young star-forming cluster do not all form at the same time. 
For instance, star formation in the star-forming region of Taurus is thought to have been ongoing for 1-2\,Myr \citep{1995ApJS..101..117K}. Across Orion, for discs in isolation, the spread in disc mass is likely related to an age spread \citep{2022A&A...661A..53V}. Therefore, for the internal age spread of the discs, we used a value of 1\,Myr. 

When a protoplanetary disc forms, it will inherit the dust-to-gas ratio of its parent molecular cloud core. The metallicity of a clump in a giant molecular cloud which forms s stellar cluster is thought to be very homogeneous, due to a very short mixing timescale \citep{2014Natur.513..523F}. Hence, we did not vary the initial dust-to-gas ratio of the cloud cores and kept the nominal value of $Z = 0.01$ for all cloud cores in the population synthesis.

Finally, we ran two population synthesis copies with the same cloud core masses and centrifugal radii, but one using the \citetalias{2021MNRAS.508.3611P} prescription and the other using the \citetalias{2021ApJ...910...51K} prescription. The discs are evolved for 10\,Myr using $\alpha_\nu = 10^{-2}$ (Sect. \ref{sec:res:alpha1e-2}) and $\alpha_\nu = 10^{-3}$ (Sect. \ref{sec:res:alpha1e-3}) for disc lifetimes of 2.5\,Myr and 8\,Myr \citep{2009AIPC.1158....3M, 2021ApJ...921...72M}, respectively.


\section{Results}
\label{sec:results}

\subsection{Evolution of a single disc} \label{sec:res:sing_disc}

 Before presenting the population synthesis results, we show the evolution of a single protoplanetary disc to illustrate the key physical processes at play. For this purpose, we initiated a model with a molecular cloud core with a mass of $1$\,$\mathrm{M}_\odot$ and a maximum centrifugal radius of $10$\,au. We used an $\alpha$-viscosity of $\alpha_\nu = 10^{-2}$, together with the \citetalias{2021MNRAS.508.3611P}-photoevaporation prescription. Other model parameters are the same as described in Sect. \ref{sec:Model:popsynth}.

\begin{figure*}[t]
\centering
\includegraphics[width=\hsize]{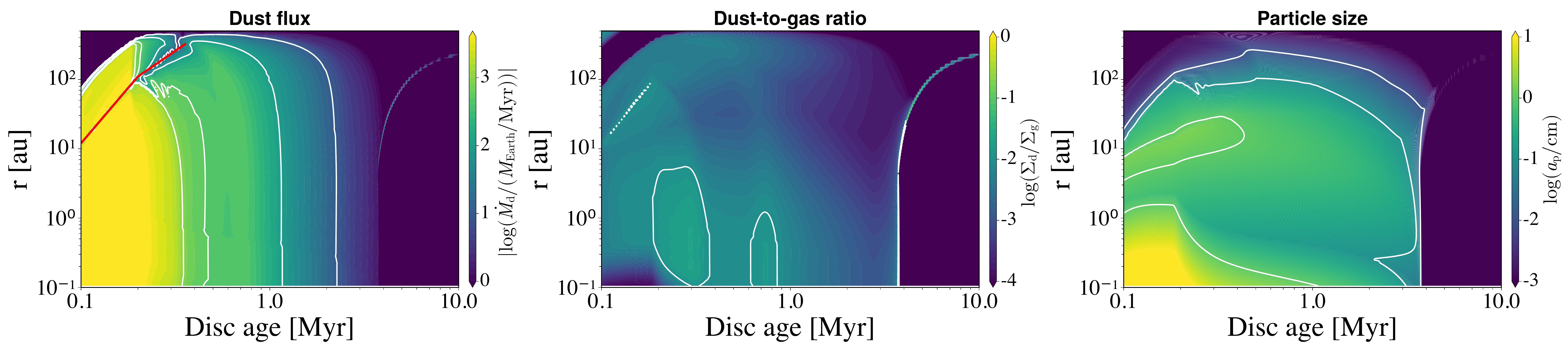}
      \caption{Dust flux (left, absolute values), dust-to-gas ratio (middle), and dust grain size (right) evolution of a disc forming from a 1 $\mathrm{M}_\odot$ cloud core with a centrifugal radius of 10\,au and photoevaporation according to \citetalias{2021MNRAS.508.3611P}. When the disc is young and the gas expands outwards by viscous evolution, the dust in the outer disc also moves outwards as it is strongly coupled to the gas. The red line marks the radius at which the dust transitions from moving inwards to moving outwards in the disc. Once photoevaporation opens a gap in the gas, the dust interior to this line is quickly drained. The dust remaining outside the gap is slowly pushed outwards as the gas at the outside edge of the gap photoevaporates. Two islands of a dust-to-gas ratio higher than the initial value of 0.01 appear. One at 0.2-0.4\,Myr at radii between 0.1-5\,au, and one at 0.6-0.9\,Myr at radii from 0.1-1.5\,au. The particle size generally is larger in the inner regions of the disc and at earlier times.}
         \label{Fig:res:contour_plots_nom}
\end{figure*}

Viscous evolution dominates the gas disc for the first $\sim$\,3.7\,Myr, after which photoevaporation begins to quickly carve out a gap in the gas. A gap in the disc forms at an orbital distance of about 3--4\,au. This is seen in Fig. \ref{Fig:res:PPD_plots_nom}, which shows the evolution of the temperature (left panel), gas surface density (middle panel), and dust surface density (right panel). Once the gap opens up, dust drift over this gap is prevented because the decreased pressure gradient at the gap edge (Fig.\,\ref{Fig:res:PPD_plots_nom}, panel b). For this nominal case, we find that 0.6\,M$_\mathrm{Earth}$ of dust was trapped outside the gap, which gradually is pushed outwards as the inner cavity expands.

The radially inwards flux of pebbles is highest in the inner disc, early in its evolution (Fig.\,\ref{Fig:res:contour_plots_nom}, left panel). However, during the early times ($\lesssim$ 0.35\,Myr) dust also has an outwards motion in the outer disc region between 10--400\,au, because the gas in this region is expanding viscously outwards and the dust is well-coupled to it. The dust flux drops over time until a gap is opened up by photoevaporation. The dust remaining within the gap is quickly pushed inwards, whilst the dust retained outside the gap is slowly pushed outwards. This latter effect is represented by the thin sliver of dust shown in green in the top right of the left panel in Fig. \ref{Fig:res:contour_plots_nom}. 

The dust-to-gas ratio exceeds the initial value of 0.01 at two regions in time and space (Fig.\,\ref{Fig:res:contour_plots_nom}, middle panel). It occurs when the disc is 0.2--0.4\,Myr old at radii between 0.1--5\,au, and at a later time, between 0.6--0.9\,Myr, at radii between 0.1--1.5\,au. The high dust-to-gas ratio makes these potential sites for planetesimal formation \citep{2017A&A...606A..80Y, 2020A&A...638A.156A, 2021ApJ...919..107L} (see also Sect.\, \ref{sec:dis:planform}). 

Particles do not grow beyond cm in size, except at early times, within $\lesssim 1$\,Myr (Fig.\,\ref{Fig:res:contour_plots_nom}, right panel). Around 1\,au there is a local minimum in the particle size distribution. This region is located where the disc transitions from the maximum temperature of 1500\,K to lower temperatures at wider radii, where stellar irradiation is dominant in setting the disc temperature rather than viscous heating. 
 
\subsection{Population synthesis with $\alpha_\nu = 10^{-2}$} \label{sec:res:alpha1e-2}

In this section, we present the results from the population synthesis model using $\alpha_\nu = 10^{-2}$. The value of $\alpha_\nu$ which best matches observations of protoplanetary discs lifetimes is still a matter of debate, as discussed in Sect.\,\ref{sec:mod:gas}. We show the results for $\alpha_\nu = 10^{-3}$ in Sect\,\ref{sec:res:alpha1e-3}.

\subsubsection{Gas disc evolution}

\begin{figure}[t]
    \centering
    \includegraphics[width=\hsize]{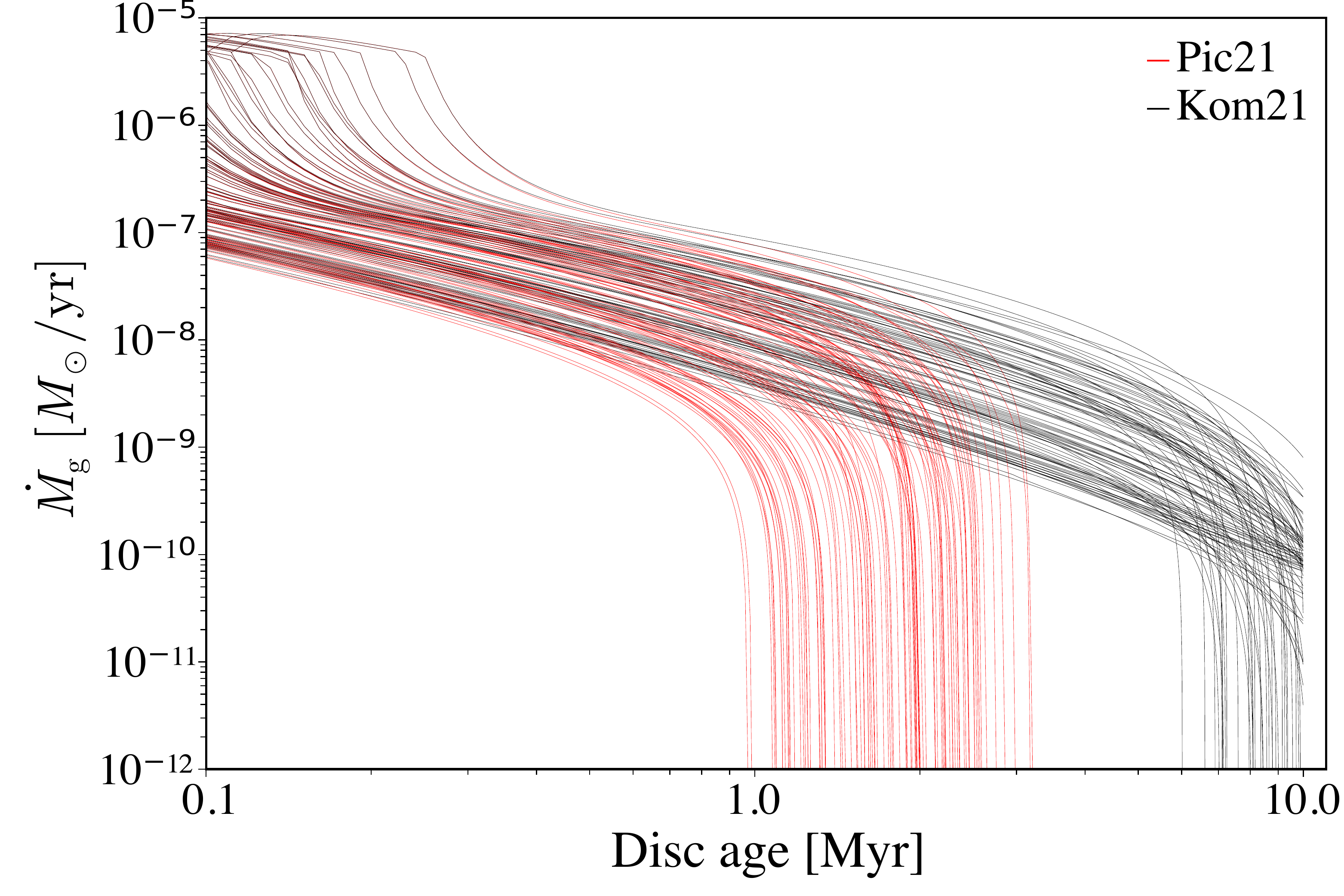}
    \caption{Gas accretion rate as a function of disc age for all disc in both photoevaporation prescriptions. Once photoevaporation opens up a gap in the gas disc, the gas within disc gap is quickly cleared by photoevaporation and viscous evolution. This results in a rapid reduction in the gas accretion rate. In this figure, we show the results without including the spread in the time of formation of the stars in a cluster (see Sect. \ref{sec:Model:popsynth}).}
    \label{fig:res:mdot_t}
\end{figure}

The evolution of the gas accretion rate over time for all discs in the population synthesis is shown in Fig. \ref{fig:res:mdot_t}.  
Discs evolving under \citetalias{2021MNRAS.508.3611P} photoevaporation (red curves) are dominated by viscous evolution for about 1--3\,Myr before photoevaporation takes over. In the \citetalias{2021ApJ...910...51K} prescription (black curves) discs undergo viscous evolution for longer times. Photoevaporation clears the inner parts of discs at the earliest after about 5--6\,Myr and for some discs, photoevaporation never plays an important role during the 10\,Myr that we simulated in this work. 

\begin{figure*}[t]
    \centering
    \includegraphics[width=\hsize]{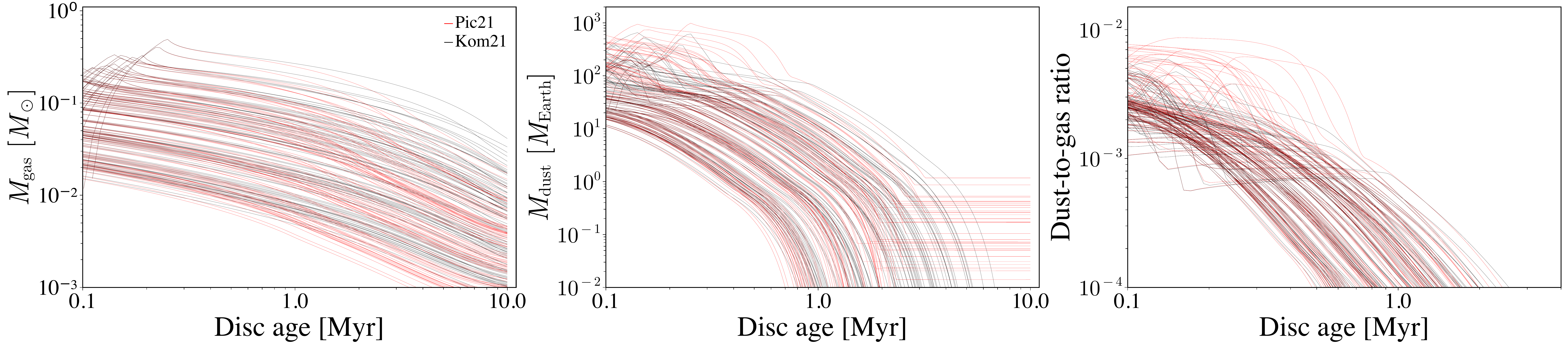}
    \caption{Evolution of the gas disc masses (left), dust disc masses (middle), and dust-to-gas ratios (right) as a function of disc age using the $\alpha_\nu = 10^{-2}$. Gas discs are formed with masses between 0.03\,M$_\odot$ to 0.5\,M$_\odot$. After 10\,Myr of evolution about 1\,\% of this initial mass remains. Dust disc masses are initially between 10\,M$_\Earth$ to 1000\,M$_\Earth$. The shortest lived dust discs are depleted in $\sim 1$\,Myr, whilst the longest lived last up to 3-8\,Myr, depending on if either the \citetalias{2021MNRAS.508.3611P} or \citetalias{2021ApJ...910...51K} photoevaporation prescription is used. Discs are formed from material with a dust-to-gas ratio of $Z=0.01$, but, due to rapid loss of dust in the earliest phases of disc formation, the global dust-to-gas ratio of the discs is always below this value.}
    \label{fig:res:Mgas_Mdust_dtg_t}
\end{figure*}

Newly formed discs have gas masses ranging from 0.03\,M$_\odot$ to 0.5\,M$_\odot$. These gas discs are then depleted by accretion and photoevaporation (Fig.\, \ref{fig:res:Mgas_Mdust_dtg_t}). For both the \citetalias{2021ApJ...910...51K} and \citetalias{2021MNRAS.508.3611P} photoevaporation prescriptions, the mass reservoir diminishes until approximately 1\,\% of the initial mass remains. This remainder of gas is found outside the photoevaporative gap at large orbital radii. Therefore, it takes excessively long to deplete by photoevaporation.

\subsubsection{Dust disc evolution} \label{sec:res:popsynth:dust}

\begin{figure*}[t]
\centering
\includegraphics[width=\hsize]{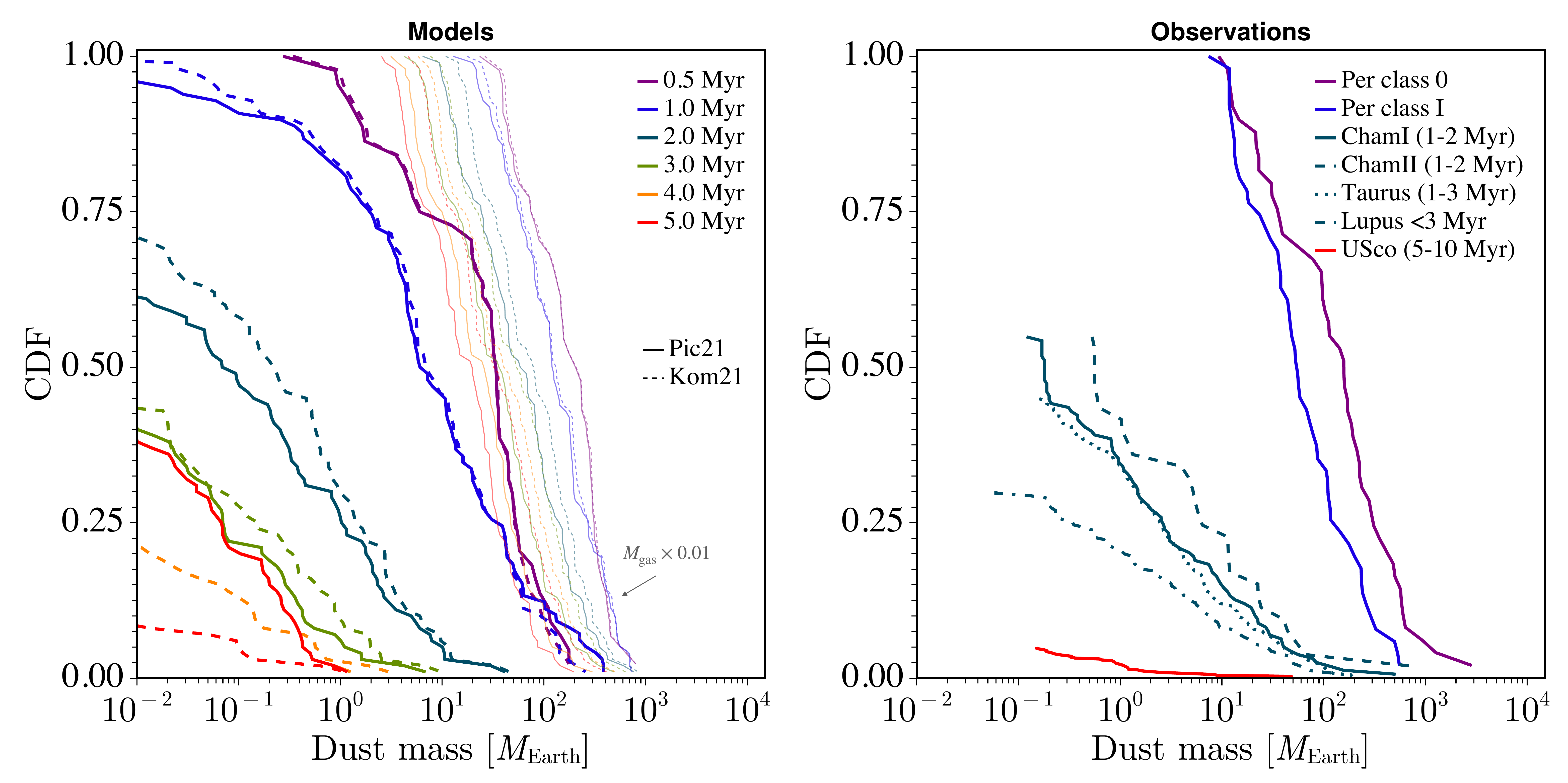}
      \caption{Cumulative disc dust mass distributions for the model (left plot) using using $\alpha = 10^{-2}$. Here we show the \citetalias{2021MNRAS.508.3611P} (solid) and the \citetalias{2021ApJ...910...51K} (dashed) photoevaporation prescriptions. A sample of observed disc masses are shown in the right plot. The two photoevaporation prescriptions are very similar at 0.5 and 1\,Myr and begin to differ at the low-mass end at 2\,Myr. At 4-5\,Myr the \citetalias{2021MNRAS.508.3611P} retains dust in some discs, whereas the \citetalias{2021ApJ...910...51K} has very little left. We note that the 3, 4, and 5\,Myr lines of the \citetalias{2021MNRAS.508.3611P} are on top of each other. The thin coloured lines in the left plot show the gas mass multiplied by the initial dust-to-gas ratio. This shows how the disc dust mass would evolve without any radial drift. The observed sample is colour coded according to the cluster's oldest inferred age.}
         \label{Fig:res:CDF_Pic+Kom_alpha1e-2}
\end{figure*}

The radial drift of pebbles reduces the dust mass of protoplanetary discs on a timescale of a few Myr. This is illustrated in the middle panel of Fig. \ref{fig:res:Mgas_Mdust_dtg_t}, which shows the evolution of the dust mass as a function of disc age. Discs are formed from cloud cores with gas masses between $\sim 0.1 \ \mathrm{M}_\odot$ and $\sim 1 \ \mathrm{M}_\odot$, and are born with dust disc masses between of 50\,M$_\Earth$ and 1000\,M$_\Earth$ respectively. The least massive and shortest lived discs are depleted of dust within less than 1\,Myr, whilst the most massive and longest lived discs last up to 7\,Myr. This process is also seen in the evolution of the cumulative distribution of dust disc masses at cluster ages between 0.5\,Myr and 5\,Myr (as shown in Fig. \ref{Fig:res:CDF_Pic+Kom_alpha1e-2}). In the 0.5\,Myr-old cluster, 73\,\% of discs have a disc dust mass higher than 10\,M$_\Earth$. At 1\,Myr, this has decreased to only 45\,\% of discs having over 10\,M$_\Earth$ of dust. From here on, we see a continued decrease in the available disc dust mass as more and more dust drifts onto the star.

We do not find that photoevaporation is able to retain significant amounts of dust by gap opening. Using the \citetalias{2021ApJ...910...51K} prescription, all discs retain  $<0.1$\,M$_\Earth$ of dust. With the \citep{2021MNRAS.508.3611P} model, we find that 20 \% of all discs are able to retain at least 0.1\,M$_\Earth$ of dust. However, among these discs, the median mass retained is only 0.3\,M$_\Earth$. The median mass of these discs right at the end of their formation, which is when they are the most massive, is 367\,M$_\Earth$. Hence, these discs retain less than 0.01\,\% of their maximum dust mass.  For comparison, the observed median dust masses in protoplanetary disc surveys at $\sim 2$\,Myr are 3.0\,M$_\Earth$ (Lupus), 3.2\,M$_\Earth$ (Cham I), 6.2\,M$_\Earth$ (Taurus), and 11.5\,M$_\Earth$ (Cham II) \citep{2013ApJ...771..129A, 2016ApJ...828...46A, 2016ApJ...816...25P, 2021A&A...653A..46V}.

\begin{figure}
\centering
\includegraphics[width=\hsize]{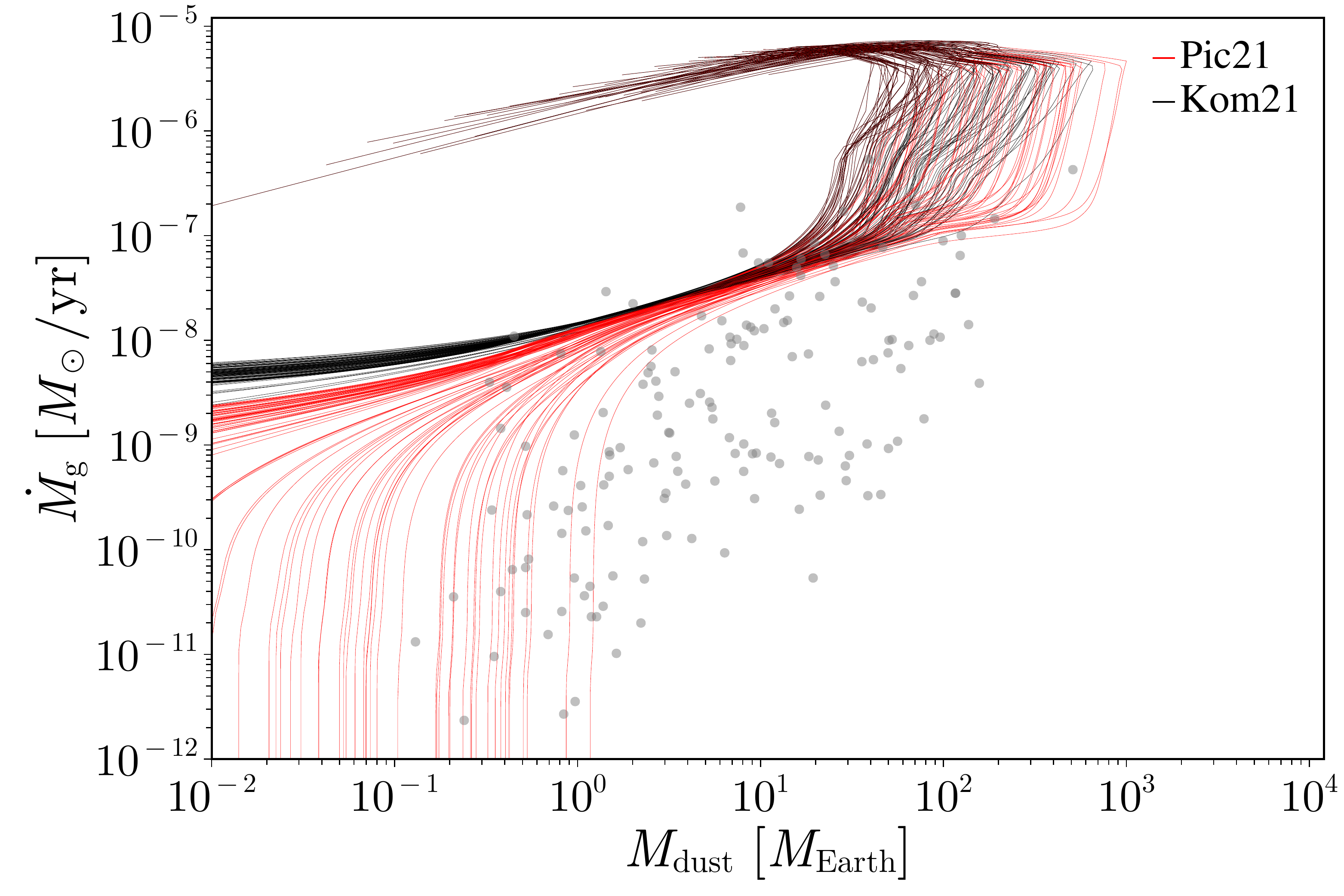}
      \caption{Gas accretion rate as a function of dust disc mass for the models using the \citetalias{2021MNRAS.508.3611P} (black) and the \citetalias{2021ApJ...910...51K} (red) photoevaporation prescriptions . The grey dots show the observational sample from \cite{2022arXiv220309930M}. Each line represents one disc's evolutionary track. Discs start out at low dust masses and high gas accretion rates ($\sim 10^{-6}\ \mathrm{M}_\odot/ \mathrm{yr}$). As they leave the formation phase, they pass through a phase where the gas accretion rate drops rapidly and the dust mass changes slowly. After this, the disc dust mass begins to quickly drain. If photoevaporation opens up a gap in the gas and clears the inner gas disc during this phase, the gas accretion rate drops rapidly and some dust is trapped. Otherwise, the gas accretion rate remains nearly unchanged while the dust drains.}
         \label{Fig:res:Mdot-Mdust_comp_alpha_1e-2}
\end{figure}

The retention of dust by gap opening can also be seen in Fig.\,\ref{Fig:res:Mdot-Mdust_comp_alpha_1e-2}, where the gas accretion rate as a function of dust disc mass is shown for each disc using both photoevaporation models. Discs start out with high gas accretion rates ($\sim 10^{-6} \ \mathrm{M}_\odot/\mathrm{yr}$) and low disc dust masses. For the  \citetalias{2021MNRAS.508.3611P} model (red curves) some of these discs reach a phase where the gas accretion turns over at rates between $ 10^{-8}$ and $10^{-9} \ \mathrm{M}_\odot/\mathrm{yr}$, due to photoevaporation clearing the inner disc of gas, whilst retaining some dust. These are the discs where photoevaporation has trapped dust outside the gas gap. None of the discs using the \citetalias{2021ApJ...910...51K} model (black lines) go through this phase, showing that all these discs lost all of their dust before photoevaporation opened up a gap.

The efficient depletion of dust by radial drift agrees with our previous results from \cite{2020A&A...638A.156A} where radial drift is the dominant process for removing dust from protoplanetary discs. Other mechanisms of dust removal, such as planetesimal formation and subsequent planet formation, would only further deplete the dust reservoir (further discussed in Sect. \ref{sec:dis:planform}). 

If radial drift would not be effective, the dust would evolve by advecting with the gas and therefore be lost at the same rate as the gas. The cumulative distribution of disc dust masses in this case is shown with the thin coloured lines in the left plot of Fig. \ref{Fig:res:CDF_Pic+Kom_alpha1e-2}. For this experiment, we calculated these dust masses by multiplying the gas mass of each disc with the initial dust-to-gas ratio and kept it fixed at a constant value of $Z=0.01$ in our simulations (Sect.  \ref{sec:Model:popsynth}). In this case, the dust is lost very slowly and the profile of the cumulative distribution remains similar throughout the evolution of the cluster. 

For comparison, we show a sample of observed dust disc masses in different star-forming regions in the right plot of Fig. \ref{Fig:res:CDF_Pic+Kom_alpha1e-2}. The observed sample is taken from \cite{2016ApJ...828...46A, 2016ApJ...827..142B, 2016ApJ...831..125P, 2020A&A...640A..19T, 2021A&A...653A..46V, 2022arXiv220309930M}. To get the occurrence fraction with respect to the full population of stars in the cluster, we scaled the cumulative distributions of this sample according to Eq. \ref{eq:disc_frac} (for more, see Appendix \ref{app:discfrac}), assuming a gas disc lifetime of 2.5\,Myr \citep{2009AIPC.1158....3M}. We note that the model disc masses presented in this paper are the masses as given directly by the model, that is the integrated surface densities. We do not take into account optical depth effects of the dust to model the dust disc mass that would effectively be observable. The observable dust disc mass would be expected to be somewhat lower than the model masses.

At a cluster age of 0.5\,Myr and 1\,Myr, the median dust disc mass of the model is $\sim 30\ \mathrm{M}_\Earth$ (0.5\,Myr) and  $\sim 5.5\ \mathrm{M}_\Earth$ (1\,Myr). This can be compared to the Perseus sample, which has significantly higher median masses for class 0 objects ($158 \ \mathrm{M}_\Earth$) and class I objects ($52 \ \mathrm{M}_\Earth$) \citep{2020A&A...640A..19T}. In these early stages of disc evolution, radial dust drift could be less efficient than we find here. However, determining the masses of young embedded discs remains challenging \citep{2018ApJ...867...43T}. 

For the more evolved discs we find that the distribution of observed discs aged between 1--3\,Myr all lie in between the 1--3\,Myr old cluster in our population synthesis. The observed disc dust masses lie closer to the 2--3\,Myr older model clusters than the 1\,Myr model clusters. For the oldest observed cluster, Upper Scorpius, only the oldest model cluster (5\,Myr) with the \citetalias{2021ApJ...910...51K} photoevaporation model lies somewhat close. Using the \citetalias{2021MNRAS.508.3611P} photoevaporation prescription, too much dust is retained in too many discs. However, the fate of solids concentrated at the cavity edge is uncertain (see Sects. \ref{sec:diss:dustoutgap} and \ref{sec:diss:photovap}). 

If we compare the observed sample to the expected dust mass for models without radial drift (thin coloured lines in Fig. \ref{Fig:res:CDF_Pic+Kom_alpha1e-2}) it is clear that the match is very poor, with the exception of the Perseus discs. This indicates that dust in discs that are a few Myr old has been lost at a higher rate than the gas. Therefore, the sometimes-used assumption that the gas mass in protoplanetary discs is 100 times their dust mass does not hold. We see this in the right panel of Fig. \ref{fig:res:Mgas_Mdust_dtg_t}, which shows the evolution of the global dust-to-gas ratio of all discs as a function of their age. We note that, even though the cloud cores from which the discs form initially have dust-to-gas ratios of 0.01, efficient radial drift of dust very early in the disc evolution decreases the global dust-to-gas ratio already very early on in the disc's evolution ($\lesssim 10^5$\,yr). 

Although radial drift is able to qualitatively match the cumulative disc dust mass distributions well, there is one component of the observational sample which is not well reproduced: the observed clusters show that there is a fraction ($\sim\,10\% $) of discs that retain large amounts ($\geq 10 \mathrm{M}_\Earth$) of dust at late times \citep{2021AJ....162...28V}. In Appendix \ref{app:halt_drift} we show an experiment of halting dust drift in a subset of the most massive discs. We find that stopping dust drift in a fraction as low as 5\,\% of the discs can extend the high-mass tail of the CDF, bringing the model more in line with the observations. This supports the claim that $\sim\,5\%-10\%$ of discs have a reduced efficiency, or complete halting, of radial dust drift. This group of discs can represent a population of discs in which radial drift is less efficient. A likely mechanism for preventing radial drift of pebbles is the presence of pressure bumps in the protoplanetary disc, possibly triggered by wide-orbit planets. We discuss this further in Sect. \ref{sec:dis:pressbump}.

An important measure of disc evolution is the relation between the gas accretion rate onto the star and the dust disc mass. This is shown in Fig.\,\ref{Fig:res:Mdot-Mdust_comp_alpha_1e-2}, where the lines shown the model and the grey dots show observational measurements from \citep{2022arXiv220309930M}. Here, we see that a large region of parameter space with relatively low accretion rates of less than about $10^{-8}\ \mathrm{M}_\odot$/yr and dust disc masses of over 1\,M$_\Earth$ is populated by the observed sample. Our synthesis model with the \citetalias{2021MNRAS.508.3611P} photoevaporation does not produce such dust-rich discs with low gas accretion rates. This region contains roughly half of the of observed sample. Using the \citetalias{2021ApJ...910...51K} photoevaporation prescription, an even smaller region of the observed sample is reached. Only those discs with accretion rates of about $10^{-8}\ \mathrm{M}_\odot$/yr or more are reproduced.

The disagreement between model and observation in the gas accretion rate and dust disc relation could originate from an overestimation in the model of the gas accretion rate onto the host star. Lower gas accretion rates onto the star are possible if disc winds remove some gas from the inner disc before it is accreted onto the host star \citep{2022arXiv220309930M}. Alternatively, dust could have been consistently retained in heavily gas-depleted discs, in apparent conflict with the evolution of the cumulative mass distribution (Fig.\,\ref{Fig:res:CDF_Pic+Kom_alpha1e-2}). Finally, our models could be consistent with the observations if the estimated mass flux onto the host star is somehow underestimated from the accretion observations.

\begin{figure*}[t]
\centering
\includegraphics[width=\hsize]{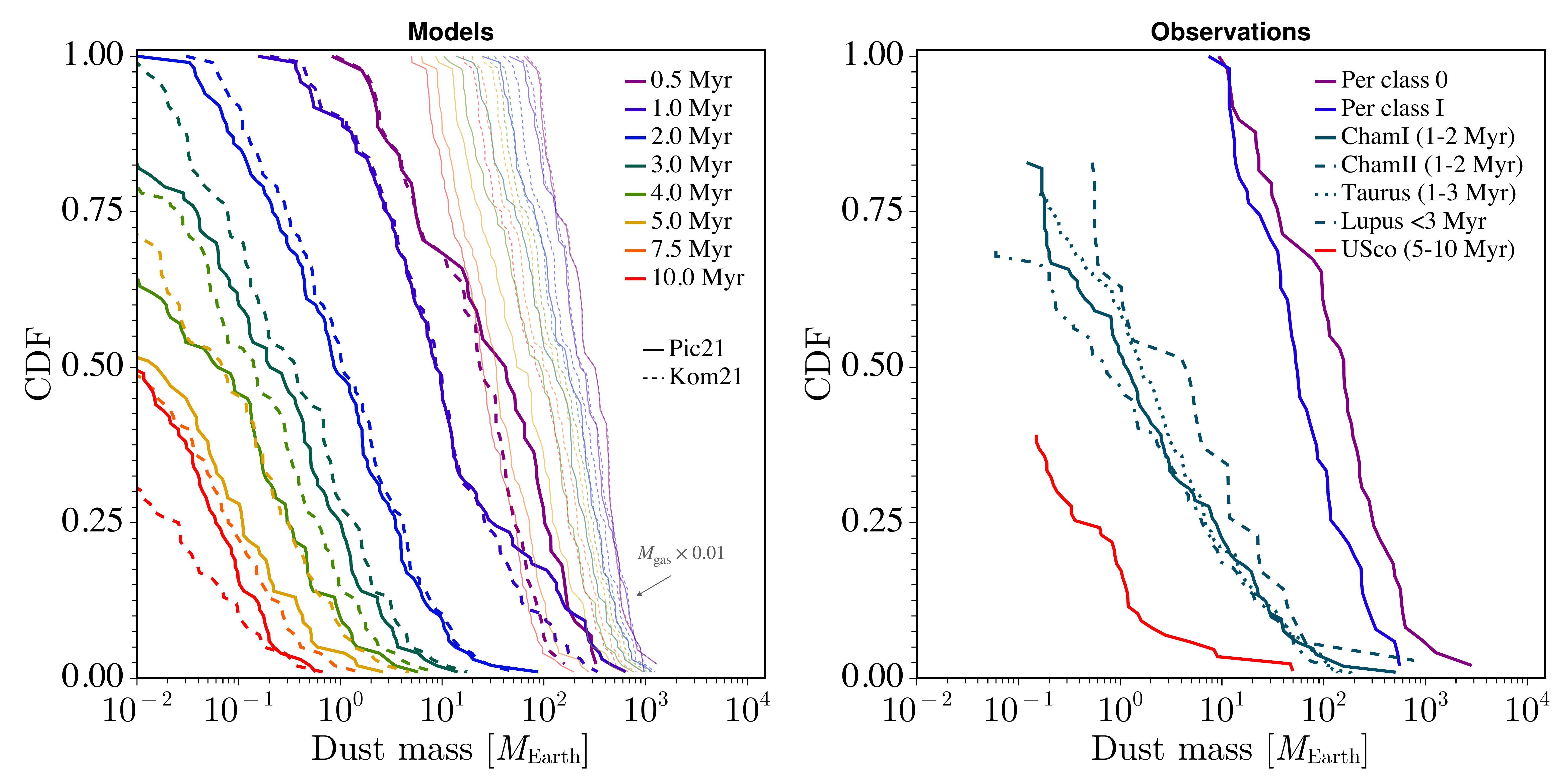}
      \caption{Cumulative distributions of disc dust masses from the population synthesis model using $\alpha = 10^{-3}$ (left) and observed discs (right). The observed sample is taken from \cite{2020A&A...640A..19T} (Perseus) and \cite{2022arXiv220309930M}. The distributions of the observed sample are scaled according to the disc fractions from \cite{2021ApJ...921...72M}, taking age of each region as the mid-point of the given age range.}
         \label{Fig:res:CDF_Pic+Kom_alpha1e-3}
\end{figure*}

\subsection{Population synthesis with $\alpha_\nu = 10^{-3}$} \label{sec:res:alpha1e-3}

Lowering the viscous $\alpha_\nu$ has a number of effects on the evolution of a protoplanetary disc. In general, the disc will evolve more slowly as angular momentum transport is less efficient. However, when the disc is very young, the increased viscosity due to gravitational instabilities can dominate over the base viscosity, making early evolution somewhat similar for different base $\alpha_\nu$ values (Eq. \ref{eq:alpha_nu}). The slower viscous evolution results in photoevaporation becoming dominant at later times and therefore longer gas disc lifetimes. Viscous heating is also less efficient, leading to a cooler disc, further slowing down the evolution.

The effect on the dust component is less straightforward. A slower gas evolution results in generally higher gas surface densities. This, in turn, results in lower Stokes numbers and slower dust drift (Eq.\, \ref{eq:dustdrift} and \ref{eq:stokes}). However, the particle size in the fragmentation limit is linearly dependant on the gas surface density (Eq. \, \ref{eq:a_frag}). Therefore, Stokes numbers do not change significantly (for a fixed value of $\alpha_\mathrm{t}$). However, the dust is also advected with the gas. At $\alpha_\nu = 10^{-3}$ the radial gas velocity is slower than at $\alpha_\nu = 10^{-2}$, resulting in less dust being removed from the disc by pure advection with the gas. The dust disc is therefore effectively drained at a lower rate at $\alpha_\nu = 10^{-3}$ compared to $\alpha_\nu = 10^{-2}$. 

We do not find that the retention of dust by photoevaporative gaps is significantly different at $\alpha_\nu = 10^{-3}$ compared to $\alpha_\nu = 10^{-2}$ (Fig. \ref{Fig:res:CDF_Pic+Kom_alpha1e-3}). The dust discs are longer lived by a few million years at $\alpha_\nu = 10^{-3}$ compared to $\alpha_\nu = 10^{-2}$, the gas discs also evolve slower and a photoevaporative gap opens up typically $1-2$\,Myr later. The fraction of discs which retain $>0.1 \ \mathrm{M}_\Earth$ of dust outside the photoevaporative gap is 20\% at $\alpha_\nu = 10^{-2}$ and  15\% at $\alpha_\nu = 10^{-3}$.  Using $\alpha_\nu = 10^{-2}$ the median dust mass retained outside the photoevaporative gap is $0.3\,\mathrm{M}_\Earth$, compared to $0.2\,\mathrm{M}_\Earth$ when using  $\alpha_\nu = 10^{-3}$.

The observed cumulative distributions (right panel of Fig.\,\ref{Fig:res:CDF_Pic+Kom_alpha1e-3}) were (as before) scaled according to Eq. \ref{eq:disc_frac}, but with a disc lifetime of $\tau = 8$\,Myr, following \citep{2021ApJ...921...72M}. This lifetime is derived by excluding regions with high external photoevaporation. It should therefore be more representative of the observational sample we are comparing to, which are not strongly exposed to external photoevaporation \citep{2021ApJ...921...72M}. 

For $\alpha_\nu = 10^{-3}$ and $\tau = 8$\,Myr, radial drift is overall consistent with the observed depletion of dust, as seen in the two panels of Fig. \ref{Fig:res:CDF_Pic+Kom_alpha1e-3}. Also, here we find that radial drift alone is not able to reproduce a small fraction of the population of discs in the observed sample, which retain large amounts of dust for more than several Myr -- similar to what we found when assuming $\alpha_\nu = 10^{-2}$. This further supports the idea that there are some discs where radial drift may be less efficient.

\begin{figure}[t]
\centering
\includegraphics[width=\hsize]{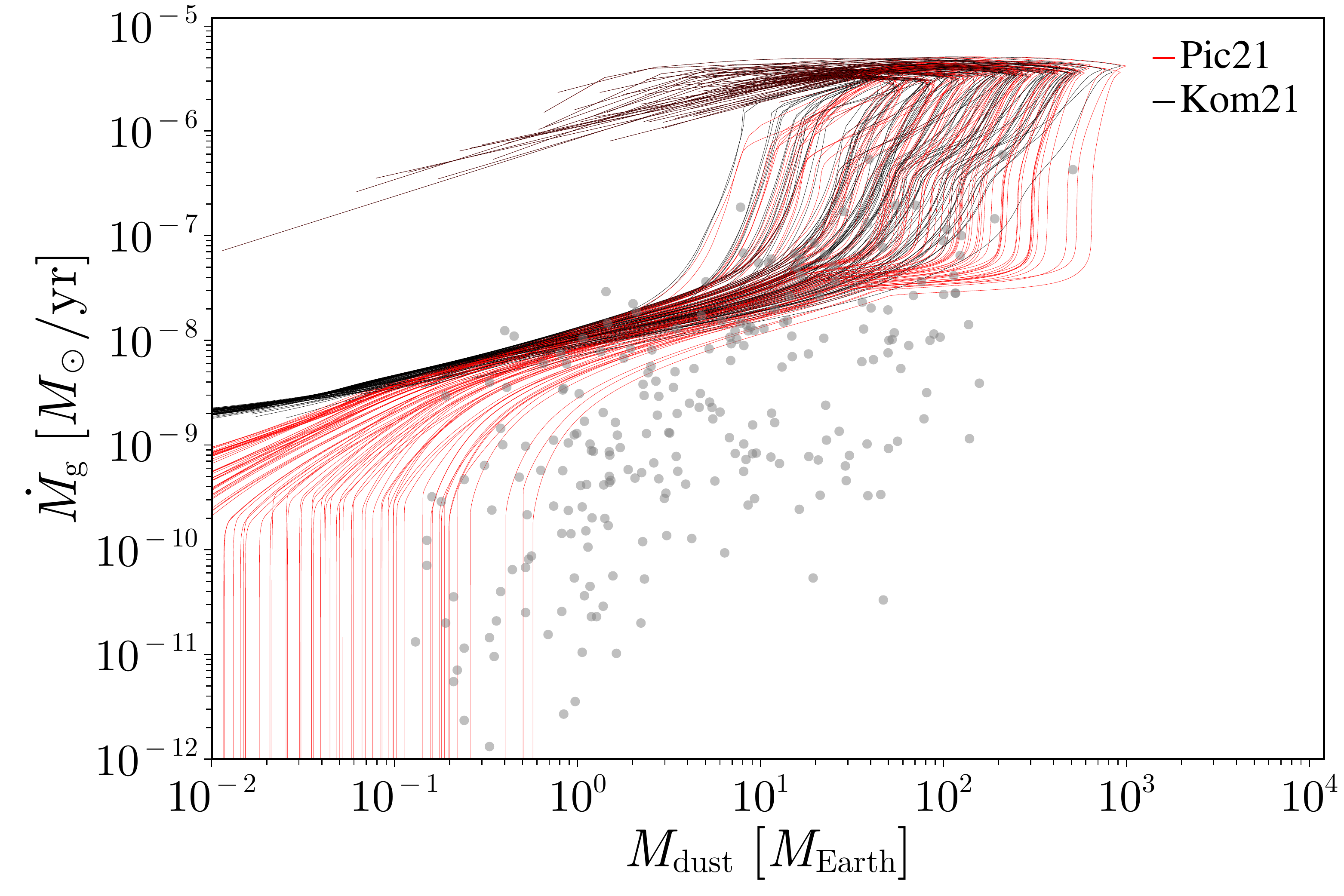}
      \caption{Gas accretion rate, assuming $\alpha_\nu = 10^{-3}$, as a function of dust disc mass for the models using the \citetalias{2021MNRAS.508.3611P} photoevaporation prescription (black) and the \citetalias{2021ApJ...910...51K} photoevaporation prescription (red).}
         \label{Fig:res:Mdot-Mdust_comp_alpha_1e-3}
\end{figure}

More slowly evolving discs lead to lower accretion rates onto host stars, which are manifested in the $\dot{M}_\mathrm{g} - M_\mathrm{dust}$ relation. However, we find that the combined evolution of both quantities is scarcely different from the $\alpha_\nu =10^{-2}$ case. There is a slightly larger scatter in gas accretion rates and a larger scatter in the disc dust mass at the end of the disc formation, as shown in Fig. \ref{Fig:res:Mdot-Mdust_comp_alpha_1e-3}.

\section{Discussion}
\label{sec:discussion}

\subsection{Interpretation of disc dust masses} \label{sec:dis:dustmass}

\subsubsection{Dependency on the stellar mass}

\begin{figure}[t]
\centering
\includegraphics[width=\hsize]{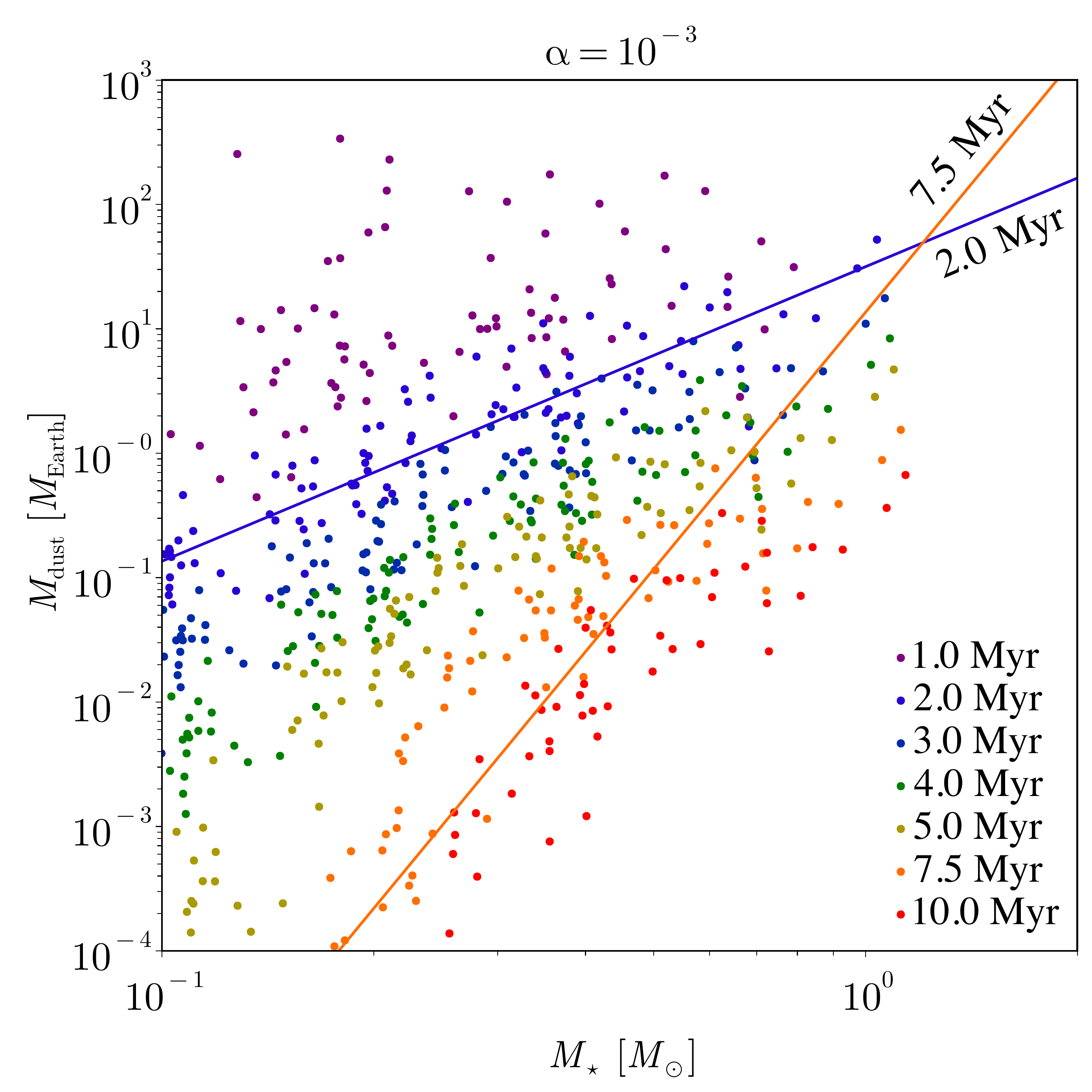}
      \caption{Dust mass as a function of stellar mass for the population synthesis model using $\alpha_\nu = 10^{-3}$ and the \citetalias{2021ApJ...910...51K} photoevaporation prescription. In discs  around lower mass stars, the dust component is removed faster than in discs around higher mass stars.}
      \label{Fig:diss:mdust_mstar_Kom_alpha1e-3}
\end{figure}

The fraction of the dust disc mass removed in a given time varies with disc and stellar mass. The dependency of the solid mass budget on stellar mass, for different cluster ages, is explicitly shown in Fig. \ref{Fig:diss:mdust_mstar_Kom_alpha1e-3}. We find that dust clearing by radial drift is more efficient around low-mass stars, as also found in previous works \citep{2020A&A...638A.156A, 2020A&A...635A.105P}. Observations point towards a linear or super-linear trend of dust mass with stellar mass, possibly steepening with time \citep{2016ApJ...831..125P}. We find steeper power law profiles, given by $M_\mathrm{dust} \propto M_\star^{\beta}$ with $\beta = 2.4$ at 2\,Myr and $\beta = 6.9$ at 7.5\,Myr, although in the latter case the data is poorly described by a single power law. When  limiting, instead, the model sample to those discs with detectable masses above 0.1\,M$_\Earth$, we can recover much more moderate power law dependencies, with $\beta = 2.2$ at 2\,Myr and $\beta = 1.6$ at 7.5\,Myr, in line with the observed exponents. These relations may further be influenced by the stellar-mass dependent occurrence of giant planets that can reduce pebble drift, as recently argued in \citet{2020A&A...635A.105P, 2021AJ....162...28V}.

This dependency on stellar mass is also reflected in the evolution over time of the profile of the cumulative disc dust mass distribution (Figs. \ref{Fig:res:CDF_Pic+Kom_alpha1e-2} and \ref{Fig:res:CDF_Pic+Kom_alpha1e-3}). The cumulative distribution becomes less steep with time. If dust removal was more efficient around more massive stars with more massive discs, the profile would become steeper with time. Conversely, if the efficiency is the same across all disc and stellar masses, the profile would appear similar in shape to the gas tracer profiles in Figs. \ref{Fig:res:CDF_Pic+Kom_alpha1e-2} and \ref{Fig:res:CDF_Pic+Kom_alpha1e-3}. In summary, radial drift not only removes dust on a timescale that agrees with observations, but it is also able to match the profile of the cumulative distribution well.

\subsubsection{Role of pressure bumps} \label{sec:dis:pressbump}

Disc surveys have revealed that many of the resolved protoplanetary discs show significant substructures, with very pronounced rings and gaps in the dust \citep[e.g.][]{2018ApJ...869L..41A}. These structures are thought to originate from pressure bumps trapping dust \citep{2018ApJ...869L..46D, 2020A&A...635A.105P, 2021A&A...645A..70P}. The high occurrence of substructures in observed discs suggests that they might be common in all discs. However, by design,  such surveys often target the largest and brightest protoplanetary discs, which could be more prone to forming gap-opening giant planets \citep{2021AJ....162...28V}.

The frequent occurrence of deep gaps and rings in the dust continuum of these large and long-lived discs may signal regions where pebble drift is delayed. The link between exoplanet statistics and disc substructures was investigated in detail by \cite{2021AJ....162...28V}. They found that the increased occurrence of extended, structured discs with stellar mass is similar to the increased occurrence of giant planets with stellar mass. This supports the idea that large scale gaps in protoplanetary discs originate from giant planets. Such pressure bumps can arise when growing planetary cores reach the pebble isolation mass and carve shallow gaps that are sufficient to halt pebble drift \citep{2014A&A...572A.107L, 2016MNRAS.459.2790R, 2018A&A...612A..30B}. Other sources of pressure bumps that can slow down or halt pebble drift also exist, for example, changes in the level of turbulence at the transition from regions where magneto-hydrodynamic instabilities are active \citep{2022arXiv220309821L}. It is thus plausible that the majority of discs are radial-drift dominated, but that in $\sim5\%-15\%$ of discs, radial drift is reduced by the presence of giant planets. This may be connected to the frequently inferred presence of giant planets interior to debris discs \citep{2022A&A...659A.135P}, where the latter could have been the natural outcome of left-over pebbles clumping into planetesimals during disc clearing \citep{2017ApJ...839...16C}.

\subsubsection{Role of planet formation} \label{sec:dis:planform}

Radial drift is not the sole process by which pebbles are removed from protoplanetary discs, as we know that a fraction of them will form the planets and planetesimals that we observe in our own solar system, as well as in exoplanet systems and debris discs. In this work, we show that radial drift removes dust from protoplanetary discs on a timescale that agrees well with dust disc lifetimes measured from observations. This suggests that radial drift is the dominant dust-removal process. However, since we do not include planetesimal or planet formation in the model, we cannot dismiss their potential importance. 

Planetesimal formation could remove a substantial fraction of dust particles present in protoplanetary discs, but its efficiency, as well as how that might vary across stellar and disc mass, is currently not well understood. Planetesimals are thought to form by the streaming instability \citep{2005ApJ...620..459Y, 2007Natur.448.1022J}, which is likely to happen in regions with a super-solar dust-to-gas ratio \citep{2015A&A...579A..43C, 2017A&A...606A..80Y, 2021ApJ...919..107L}. Pressure bumps at the gap edges of giant planets are a promising location for this process, as dust piles up in these regions and large amounts of planetesimals could potentially form \citep{2020A&A...635A.110E}. Such planetesimals would then represent a second generation of late-formed planetesimals, as they are triggered by the presence of an already fully-formed giant planet.   

Similarly, the formation of planets should also be responsible for the removal of some fraction of the dust in protoplanetary discs. To assess the importance of this process, a possible future extension of this work could include a exoplanet synthesis model as well \citep[similarly to the study of e.g.][]{2019A&A...632A...7L}. In this context, it is interesting to note that the mass of solids present in typical observed exoplanet systems is comparable to that found in class II discs \citep{2021ApJ...920...66M}. In the earlier disc stages, the mass budget in pebbles is larger by more than one order of magnitude, so up until the class II stage the decrease in the mass in pebbles is mainly driven by dust drift, rather than planet formation.

\subsubsection{What happens to the dust retained outside the photoevaporative cavity} \label{sec:diss:dustoutgap}

In our model, the dust retained outside the photoevaporative cavity eventually moves into the location of the pressure maxima where there is no radial drift. Since our model does not consider any additional treatment of this dust, this results in narrow long-lived rings of dust that are slowly pushed outwards as the photoevaporative gap in the gas expands outwards. The high dust-to-gas ratio in these pressure bumps could make them favourable locations for the formation of planetesimals via the streaming instability \citep{2021AJ....161...96C, 2022ApJ...933L..10C}. However, the fate of the dust retained outside photoevaporative gaps is uncertain. For example, \cite{2019MNRAS.487.3702O} found that this dust is effectively lost in a process where the largest particles grow until they fragment by collisions, replenishing the reservoir of small dust particles that are subsequently removed by radiation pressure. If the dust is turned into planetesimals, this could lead to the formation of debris discs. However, the median mass of dust outside the photoevaporative gaps is only $0.3\ \mathrm{M}_\Earth$ for $\alpha_\nu = 10^{-2}$ and $0.2\ \mathrm{M}_\Earth$ for $\alpha_\nu = 10^{-3}$. While this dust mass is similar to that present in the known debris disc population, the planetesimal population that creates this dust in debris discs could be significantly higher \citep{2017MNRAS.470.3606H, 2021MNRAS.500..718K}.

\subsubsection{Role of stellar binarity}

Most solar mass stars reside in multiple systems, and the stellar multiplicity fraction increases with stellar mass \citep{2022arXiv220310066O}. Observations have found that discs around binaries typically host dust masses below $50$\,M$_\Earth$, and that for binary separations below 100\,au, discs have dust masses below 10\,M$_\Earth$ \citep{2020MNRAS.496.5089Z}.  Therefore, measuring the mean disc masses in young stellar clusters without taking into account the binarity of stars can lead to underestimations of the typical masses of discs around single stars. A future extension of this work could include the effect of stellar binarity in driving faster disc dispersal \citep{2012ApJ...745...19K} through tidal effects and stellar encounters truncating disc radii in the cluster stage\citep{2016MNRAS.457..313P}

\subsection{Transition discs}

Transition discs with dust-depleted inner cavities and low accretion rates are suggested to to have been caused by the photoevaporation of the inner disc \citep{2011MNRAS.412...13O, 2022arXiv221005539V}. However, we find that it is difficult to form discs that both show signs of gas accretion onto the hosts star and massive extended dust disc at wider radii. Gas within the photoevaporation gap is lost on very short timescales ($\lesssim 10^5$\,yr), as seen in Fig. \ref{fig:res:mdot_t}, while only retaining small amounts of dust trapped at the outer gap edge. 

As discussed earlier in ths paper, the fate of the retained dust is such that it might be lost on short timescales (Sect. \ref{sec:diss:dustoutgap}). Therefore we do not find that photoevaporation, as formulated here,  is a likely pathway for forming the majority of transition discs. This is in line with the low occurrence of discs with a cavity radius within 10\,au \citep{2022arXiv220408225V} that would have been expected in classic photoevaporation models \citep{2019MNRAS.487..691P}.

\subsection{Evolution of the outer gas and dust disc radius}

\begin{figure}[t]
\centering
\includegraphics[width=\hsize]{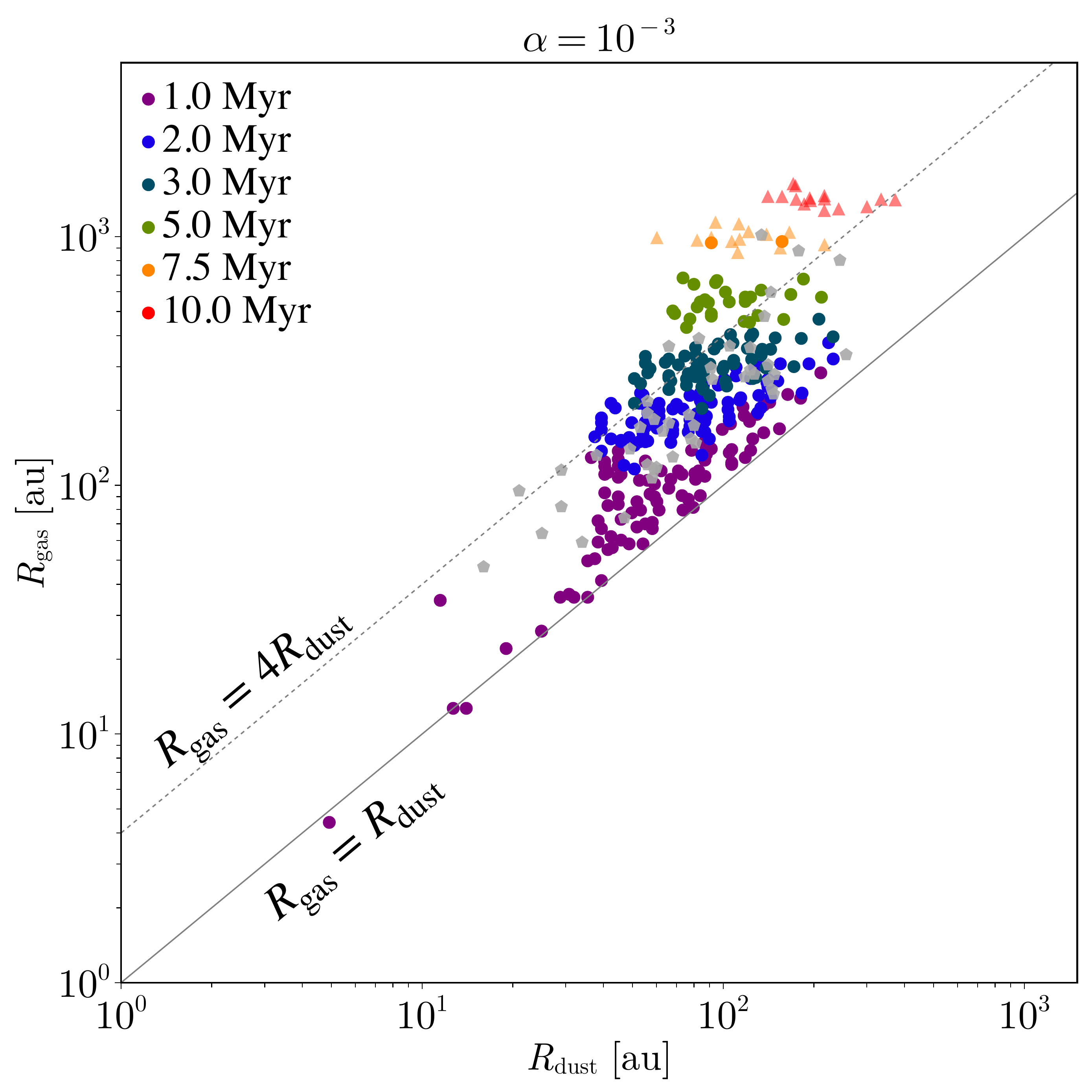}
      \caption{Gas disc radius as a function of dust disc radius, using $\alpha_\nu = 10^{-3}$ and the \citetalias{2021MNRAS.508.3611P} photoevaporation prescription for different cluster ages. Triangles show discs which have retained more than 0.1\,M$_\Earth$ of dust, but whose gas accretion rates are lower than $10^{-12} \ \mathrm{M}_\odot$/yr. These are the discs, corresponding to 15\% of the initial disc population, in which photoevaporation has opened up a gap, with a detectable amount of dust is retained outside the gap. Grey markers show observational measurements from \cite{2022ApJ...931....6L}. Grey lines show the $R_\mathrm{gas} = R_\mathrm{dust}$ (solid) and $R_\mathrm{gas} = 4 R_\mathrm{dust}$ (dashed).}
      \label{Fig:diss:rgas_rdust_alpha1e-3}
\end{figure}

Pebble drift could also, in principle, be identified through tracing the evolution of the outer dust disc radius with respect to the outer gas disc radius \citep{2019A&A...629A..79T, 2021MNRAS.507..818T}. This is nevertheless challenging as both radii could be influenced by late accretion events of gas and dust onto the disc \citep{2017ApJ...846....7K}. Moreover, in our model, the outer gas edge continuously viscously expands outwards with time, while recent models have argued against this based on disc wind models \citep{2022arXiv220309930M}. However, an outwardly expanding discs is seen in disc wind models that include the truncation of the outer disc \citep{2021ApJ...922..201Y}.

Figure \ref{Fig:diss:rgas_rdust_alpha1e-3} illustrates the evolution of the gas radius as a function of the dust radius for the population synthesis model using $\alpha_\nu = 10^{-3}$ and the \citetalias{2021MNRAS.508.3611P} photoevaporation prescription.
Here, we calculate the disc radius as the radius within which either 90\% of the disc gas or dust mass is present. If a disc has opened up a gap by photoevaporation and retained > 0.1\,M$_\Earth$ of dust outside the gap, the dust radius is taken as the outer radius where the dust is trapped and indicated with a triangle symbol. 

For very young discs, the gas and dust radii are similar because the dust is tightly coupled to the gas. As the disc evolves, the dust decouples from the gas and the dust radius begins to decrease, while the gas radius still increases. In the  approximation used here, which consists of only tracing the largest particle size, the radius of the dust disc shrinks rapidly once radial drift becomes efficient while gas radii continue to expand due to viscous evolution. Finally, the triangular symbols represent those discs where photoevaporation has opened up a gap in the gas, whilst retaining $>0.1 \ \mathrm{M}_\Earth$ of dust outside the gap. The dust disc radii also increase, as the dust is pushed outwards as the outer edge of photoevaporative gap expands.  

A direct comparison of synthetic with observed dust and gas radii is not straightforward. This would require post-processing our results with radiative transfer in order to determine the observable outer dust edge and gas edge, the latter using a CO gas tracer \citep{2021MNRAS.507..818T}. Nevertheless, for illustrative purposes, we show in Fig. \ref{Fig:diss:rgas_rdust_alpha1e-3} (in grey points) the dust and gas (CO) radii determined for the discs in the sample by \cite{2022ApJ...931....6L}. This uniformly analysed collection of discs, drawn mainly from Taurus, Lupus, Ophiuchus, and Upper Sco, spans stellar masses between 0.15–2\,M$_\odot$ and ages between 0.5 to 20\,Myr. Similarly to our results, all observed discs have dust radii located within their gas disc radii and only a few discs have severely drift-depleted discs with $R_\mathrm{gas} > 4R_\mathrm{dust}$. Previous works have argued that radial drift would rapidly deplete pebble discs and result in discs with a gas-to-dust size ratio smaller than four \citep{2021MNRAS.507..818T}. However, this is likely an effect of assuming larger fragmentation velocities ($u_\mathrm{f}=10$\,m/s), compared to the parameters used in the current study ($u_\mathrm{f}=1$\,m/s, see Sect.\,\ref{sec:dustsize}). This results in larger pebbles closer to the drift limit and an outer dust radius that moves inwards too rapidly.

\subsection{Uncertainty in photoevaporation prescriptions} \label{sec:diss:photovap}

The disagreement in derived mass-loss rates in different photoevaporation models reflects the fact that the underlying physics driving photoevaporation is not entirely understood. \cite{2022MNRAS.514..535S} found that the chosen X-ray spectrum frequency matters significantly for the derived photoevaporation rates. This can explain why some studies \cite[e.g.][]{2017ApJ...847...11W} find very low mass loss rates for for the derived photoevaporation rates X-rays. Also, X-ray photoevaporation models, such as the one from \citetalias{2021MNRAS.508.3611P} used here, may underestimate the cooling efficiency in the discs and therefore overestimate the photoevaporation rate \cite{2017ApJ...847...11W, 2022MNRAS.514..535S}.
Since we cover both low and high photoevaporation efficiencies here, we believe our conclusion on their relative small effect on dust evolution are robust. Nevertheless, this remains a clear area for future work, especially as recent works point towards photoevaporation and magnetic disc wind losses to be tightly linked \citep{2016ApJ...818..152B}.

Since our model does not track the population of small ($\lesssim 100\,\mu$m) dust grains and the one from \citetalias{2021ApJ...910...51K} provides no scaling with the dust-to-gas ratio (a ratio of 0.01 is used), we did not model the effect that a decreasing or increasing amount of small dust present has on the FUV photoevaporation rate. A decrease in the available small dust grains reduces the opacity and thus increases the penetration depth of the photons, which can enhance the mass-loss rate. Simultaneously, fewer small dust grains also leads to less photoelectric heating, which reduces the mass-loss rate. \cite{2015ApJ...804...29G} found that these two effects can to a large degree negate each other. However, at dust-to-gas ratios above $\sim 0.03$, \cite{2018ApJ...865...75N} found that a decrease in the dust-to-gas ratio increases the FUV photoevaporation rate. At lower dust-to-gas ratios, the photoevaporation rate instead decreases, but the inclusion of hydrogen ionisation by X-rays lowers the reduction in the photoevaporation rate due to the large amount of free electrons allowing for rapid recombination onto the dust grains. Similarly, \citet{2021ApJ...915...90N} also concluded that at dust-to-gas ratios below 0.01 photoevaporation by FUV is reduced due to inefficient photoelectric heating. With these results in mind, coupled with the fact that the discs in our model rarely exceed the initial dust-to-gas ratio of 0.01 (see middle panel of Fig. \ref{Fig:res:contour_plots_nom}), including the effect of small grains of the photoevaporation rate would likely result in a reduced FUV photoevaporation rate compared to our current model. Since we already find that FUV photoevaporation has a negligible effect on the dust disc evolution, our conclusions would not change.

We also did not take into account that small dust particles may also be entrained in photoevaporative winds \citep{2011MNRAS.412...13O, 2016MNRAS.461..742H, 2016MNRAS.463.2725H}, but this only affects very small particles of dust, as the maximum particle size that can be effectively entrained in photoevaporative winds is $\sim 1 \ \mu$m \citep{2021MNRAS.501.1127H, 2021MNRAS.502.1569B}. Therefore, this process does not impact significantly the mass budget of pebbles in the disc, but may be important for the determination of  gas mass loss rates in the upper parts of protoplanetary discs \cite{2017ApJ...847...11W}. 

Finally, in this project, we did not include external photoevaporation. This choice is appropriate for the well studied nearby stellar clusters considered here. Therefore, we did not include it in our model. Nevertheless, protoplanetary discs in more massive clusters with O and B stars would be subject to external photoevaporation from the strong FUV background radiation of said stars. This process may significantly shorten the disc lifetimes \citep{2022EPJP..137.1132W}.

\section{Conclusions} \label{sec:conclusions}

\begin{enumerate} 
    \item We found that the depletion of protoplanetary dust by radial drift of pebbles is compatible with the depletion trend seen in observed protoplanetary discs. We explored dust depletion in synthetic clusters discs with disc depletion timescales of 2.5\,Myr or 8\,Myr, corresponding to, respectively, a high and low viscous alpha ($\alpha_\nu = 10^{-2}$ and $\alpha_\nu = 10^{-3}$). Both cases broadly reproduce the cumulative loss off pebbles from protoplanetary discs.
    \item Our synthesis model is consistent with the evolution of the cumulative dust disc mass distribution with time. However, a fraction of about 5-10\% of observed discs show signs of reduced radial drift compared to the model, possibly triggered by wide-orbit planet formation. Other relations that are  reproduced well are the moderate decrease of dust radius with gas radius with time and the trend between disc mass and stellar mass. However, the synthesis model does not recover a population of massive dust discs with low stellar accretion rates. The latter may point to our model not including inner disc winds, which could reduce the mass accretion rate onto the host star, or the possibility that the observations underestimate stellar accretion rates.
    \item Discs forming with a global dust-to-gas ratio of 0.01 do not retain this proportion of dust. Efficient radial drift, even during the build-up of the discs, lowers the disc's global dust-to-gas ratio below that of its birth environment. Hence, gas disc masses cannot be estimated by multiplying the dust mass by the ISM-ratio of 100, even without considering dust optical depth effects.
    \item Photoevaporation plays a minor role on the evolution of the disc dust mass. Strong X-ray photoevaporation is able to retain some dust outside the photoevaporative gaps, but only a very small fraction ($\lesssim$0.01 \%) of the initial dust mass of the disc ($50-1000\ \mathrm{M}_\Earth$). Photoevaporation may aid in reaching the low gas accretion rates observed ($\lesssim 10^{-9} \ \mathrm{M}_\odot$/yr) in protoplanetary discs  \citep{2020MNRAS.498.2845S}. However, this depends on the assumed photoevaporation efficiency, and is only achieved for maximally efficient X-ray photoevaporation prescriptions.
\end{enumerate}

\begin{acknowledgements}

We thank Anders Johansen for his  comments on the manuscript. We also thank Andrew Sellek for helpful discussions and input on the photoevaporation models. We also thank the anonymous referee for their comments that helped improve this paper. J.A. acknowledges the Swedish Research Council grant (2018-04867, PI A.\,Johansen). M.L. acknowledges the Wallenberg Academy Fellow grant (2017.0287, PI A.\,Johansen) and ERC starting grant 101041466-EXODOSS.

\end{acknowledgements}

\bibliographystyle{aa}
\bibliography{References}


\begin{appendix}

\section{Photoevaporation prescriptions} \label{app:model:PE}

\subsection{\citet{2021MNRAS.508.3611P} photoevaporation} 

\citetalias{2021MNRAS.508.3611P} provides a photoevaporation prescription derived from models of X-ray photoevaporation for different stellar masses using the photoevaporation model of \cite{2019MNRAS.487..691P, 2021MNRAS.508.1675E}. 

\begin{table}[h]
\caption{Parameters of Eq. \eqref{eq:sigPE_Pic} and Eq. \eqref{eq:MdotPE_Pic} for different stellar masses.}
\label{tab:PEparams_Pic}
\begin{tabular}{l|llll}
$M_\star$ {[}$M_\odot${]} & 0.1 & 0.3 & 0.5 & 1.0 \\ \hline
a & -3.8337 & -1.3206 & -1.2320 & -0.6344 \\
b & 22.9100 & 13.0475 & 10.8505 & 6.3587 \\
c & -55.1282 & −53.6990 & −38.6939 & −26.1445 \\
d & 67.8919 & 117.6027 & 71.2489 & 56.4477 \\
e & -45.0138 & −144.3769 & −71.4279 & −67.7403 \\
f & 16.2977 & 94.7854 & 37.8707 & 43.9212 \\ 
g & −3.5426 & 26.7363 & −9.3508 & −13.2316 \\ 
\end{tabular}
\end{table}

The photoevaporation rate is given by
\begin{align}
\dot{\Sigma}_\mathrm{PE} \left( r \right) = \ln \left( 10 \right) \nonumber
\Bigg{(} &\dfrac{6 a \ln\left(r\right)^5}{r \ln \left( 10 \right)^6} 
+  \dfrac{5 b \ln\left( r \right)^4}{r \ln \left( 10 \right)^5}
+  \dfrac{4 c \ln\left( r \right)^3}{r \ln \left( 10 \right)^4} \\ \nonumber
+ &\dfrac{3 d \ln\left( r \right)^2}{r \ln \left( 10 \right)^3}
+  \dfrac{2 e \ln\left( r \right)  }{r \ln \left( 10 \right)^2}
+ \dfrac{  f}{r\ln \left( 10 \right) } \Bigg{)} \\ 
 &\dfrac{\dot{M}_\mathrm{PE}\left( r\right)}{2\pi r}\ [\mathrm{M}_\odot\ \mathrm{au}^{-2} \ \mathrm{yr}^{-1}], \label{eq:sigPE_Pic}
\end{align}
where $\dot{M}_\mathrm{PE}\left( r \right)$ is the  mass loss rate due to photoevaporation at a given disc radius. It is given by
\begin{align}
    \dfrac{\dot{M}_\mathrm{PE} \left( r \right)}{\dot{M}_\mathrm{PE}\left(L_\mathrm{X}\right)} =   10^{a\log\left(r\right)^6 +
        b\log\left(r\right)^5 + 
        c\log\left(r\right)^4 + 
        d\log\left(r\right)^3 + 
        e\log\left(r\right)^2 + 
        f\log\left(r\right)  + 
        g}, \label{eq:MdotPE_Pic}
\end{align}
where $\dot{M}_\mathrm{PE}\left(L_\mathrm{X}\right)$ is the total mass loss rate due to X-ray photoevaporation. The parameters a,..., g for different stellar masses are given by Table \ref{tab:PEparams_Pic}. The X-ray luminosities for each stellar mass is shown in Table \ref{tab:PElum_Pic}.

\begin{table}[th]
\caption{Stellar masses and corresponding X-ray luminosities used in the \citetalias{2021MNRAS.508.3611P} photoevaporation model.}
\label{tab:PElum_Pic}
\begin{tabular}{l|l}
$M_\star \ [\mathrm{M}_\odot]$ & $L_\mathrm{X}$ [erg s$^-1$] \\ \hline
0.1 & 28.8  \\
0.3 & 29.5  \\
0.5 & 29.8  \\
1.0 & 30.3  \\
\end{tabular}
\end{table}

\subsection{\citet{2021ApJ...910...51K} photoevaporation} \label{sec:model:PE}

\begin{table}[]
\caption{Parameters of Eq. \eqref{eq:sigPE_Komaki} for different stellar masses.}
\label{tab:PEparams_Kom}
\begin{tabular}{l|llllll}
$M_\star$ {[}$M_\odot${]} & 0.5 & 0.7 & 1.0 & 1.7 & 3.0 & 7.0 \\ \hline
$c_5$ & 1.06 & 0.693 & 0.131 & 1.37  & 0.33 & 0.594 \\
$c_4$ & -1.05 & -0.95 & -0.465 & -1.41 & -0.786 & -1.00 \\
$c_3$ & -0.236 & -0.038 & 0.451 & -1.42 & 0.786 & 0.234 \\
$c_2$ & 0.570 & 0.678 & 0.376 & 1.30 & 0.557 & 0.513 \\
$c_1$ & -1.62 & -1.67 & -1.67 & -1.06 & -1.58 & -1.85 \\
$c_0$ & -12.7 & -12.6 & -12.6 & -12.6 & -13.1 & -12.1
\end{tabular}
\end{table}

The photoevaporation prescription of \citetalias{2021ApJ...910...51K} is based on models which includes photoevaporation due to EUV, FUV, and X-rays. The rate of photoevaporation is given by Eq. \eqref{eq:sigPE_Komaki}:
\begin{align}
        \log\left( \dfrac{\dot{\Sigma}_\mathrm{g, PE} \left( r \right)}{ 1 \ \mathrm{g} \ \mathrm{s^{-1} \ \mathrm{cm}^{-2}} } \right) = c_5 x^5 + c_4 x^4 + c_3 x^3  + c_2 x^2 + c_1 x + c_0,  \label{eq:sigPE_Komaki}
\end{align}
where $x$ is given by
\begin{align}
         x = \log\left( \dfrac{r}{r_\mathrm{g}} \right), \label{eq:x}
\end{align}
and $r_g$ is the the gravitational radius given by:
\begin{align}
        r_\mathrm{g} = \dfrac{G M_\star}{\left( 10 \ \mathrm{km} \ \mathrm{s}^{-1} \right)^2} = 8.87 \ \mathrm{au} \left(\dfrac{M_\star}{1\mathrm{M}_\odot}\right). \label{eq:rg}
\end{align}
The photoevaporation prescription of \citetalias{2021ApJ...910...51K}    is in the range $0.1 r_\mathrm{g}\leq r \leq 20 r_\mathrm{g}$, as this is where photoevaporation is thought to occur \citep{2003PASA...20..337L}. We therefore only include photoevaporation in the same range.

The parameters $c_{5,..., 0}$ in Eq. \eqref{eq:sigPE_Komaki} for different stellar masses are given in Table \ref{tab:PEparams_Kom}. X-ray and FUV luminosities for the same stellar masses are given in Table \ref{tab:PElum_Kom}. To approximate the evolution of the photoevaporation rate as the star grows more massive we linearly interpolate between these stellar masses and parameter values.

\begin{table}[h]
\caption{Stellar masses and corresponding FUV and X-ray luminosities used in the \citetalias{2021ApJ...910...51K} photoevaporation model.}
\label{tab:PElum_Kom}
\begin{tabular}{l|llllll}
$M_\star \ [\mathrm{M}_\odot]$ & $\log L_\mathrm{FUV}$ [erg s$^-1$]  & $\log L_\mathrm{X}$ [erg s$^-1$]  \\ \hline
0.5                   & 30.9  & 29.8  \\
0.7                   & 31.3  & 30.2  \\
1.0                   & 31.7  & 30.4  \\
1.7                   & 32.3  & 30.7  \\
3.0                   & 32.9  & 28.7  \\
7.0                   & 36.5  & 30.8  
\end{tabular}
\end{table}
\vspace{-14pt}
\section{Disc fractions} \label{app:discfrac}

The dust present in protoplanetary discs will result in an excess of infrared emission at mm wavelengths. As a disc in a cluster of young stars evolves, the dust component is depleted and the infrared excess is lost. To determine the fraction of stars that have discs with mm emission,  often only stars with discs that show an infrared excess are used as targets. A number of attempts have been made at fitting a function describing how the infrared disc fraction evolves over time given a typical disc lifetime  \citep[e.g.][]{2001ApJ...553L.153H, 2009AIPC.1158....3M, 2014A&A...561A..54R, 2021ApJ...921...72M}.   

The fitting function for the disc infrared fraction often used is
\begin{align}
    f_\mathrm{disc} = e^{-t/\tau}, \label{eq:disc_frac}
\end{align}
where $\tau$ is the typical disc lifetime, and $t$ is the age of the cluster. Disc lifetimes were long thought to be a few Myr, for example, $\tau = 2.5 \ \mathrm{Myr}$ \cite{2009AIPC.1158....3M}. However, these lifetimes were derived using samples of young star clusters where external disc photoevaporation is both present and not. External photoevaporation can significantly shorten disc lifetimes. A more recent study of clusters without external photoevaporation found that discs in such environments have typical lifetimes of about 8\,Myr \citep{2021ApJ...921...72M}.

The ages of the star-forming regions in the observational sample we made our comparison to were taken from \cite{2022arXiv220309930M}. We assumed that the of a cluster age is the mid-point of the given age range. These ages are then used in Eq. \ref{eq:disc_frac} to calculate the disc fraction of a star forming region. The resulting disc fraction is then used to scale the cumulative distribution functions of dust disc masses.

\vspace{-10pt}

\section{Halting dust drift in high-mass discs} \label{app:halt_drift}

\begin{figure}[b]
\centering
\includegraphics[width=\hsize]{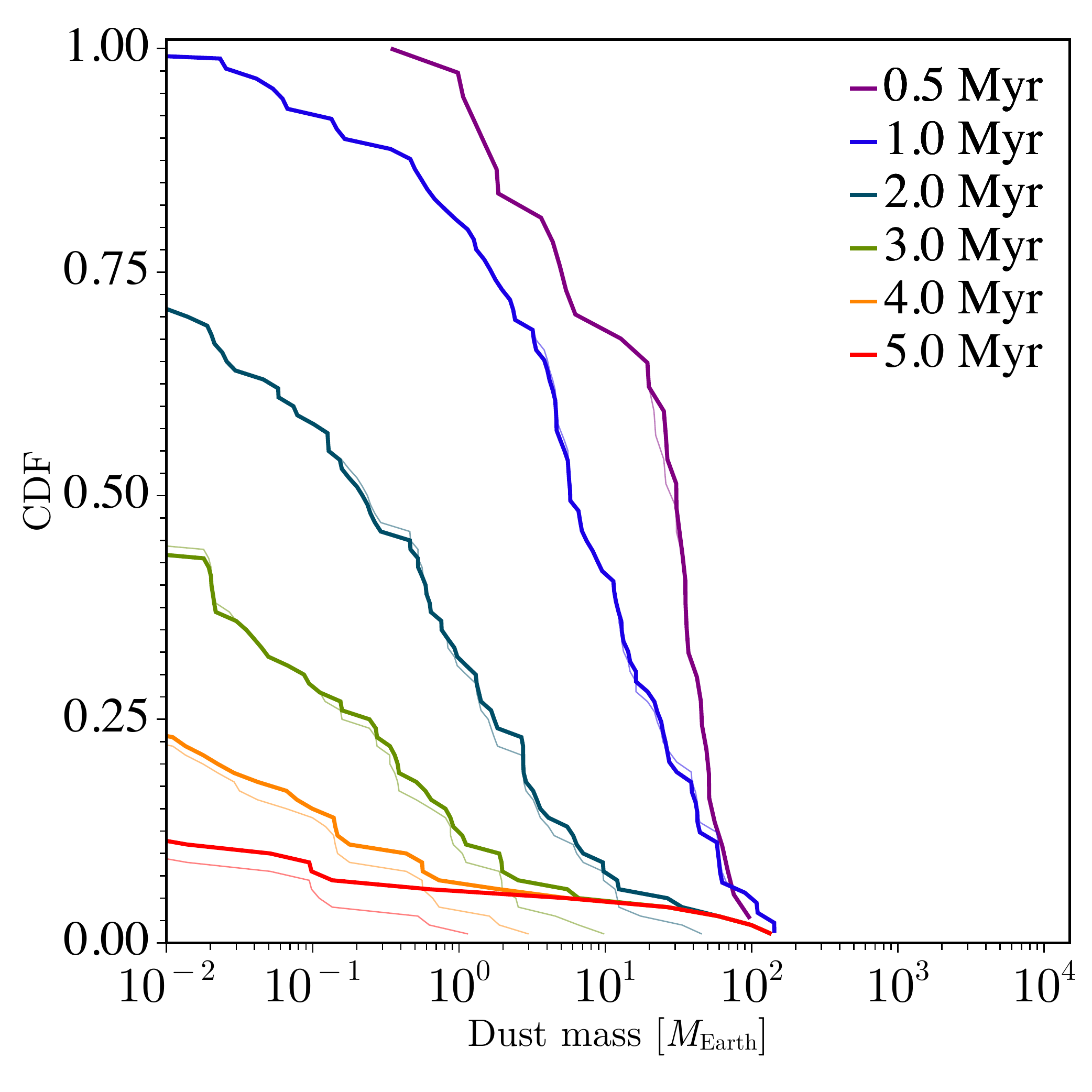}
      \caption{Cumulative dust mass distributions of the simulations using $\alpha_\nu = 10^{-2}$ and the \citetalias{2021ApJ...910...51K} photoevaporation prescription with pebble drift (thin lines) and with pebble drift halted in 5\% of discs (thick lines). Halting pebble drift extends the high-mass tail of the CDF, bringing it more in line with what is seen in observations.}
      \label{Fig:app:CDF_drift_halt}
\end{figure}

Here, we illustrate how the halting of dust transport inwards in the disc, in a fraction of the most massive discs, can affect the cumulative dust disc mass distributions. Stopping (or slowing down) the inwards transport of dust could be caused by pressure bumps in the gas disc, which are discussed in Sect. \ref{sec:dis:pressbump}. In order to approximate dust trapping in pressure bumps we halted the pebble drift between the end of the disc formation stage and 2\,Myr in 17\% of the 30\% most massive discs (corresponding to 5\% of the total disc population). We show our results in Fig. \ref{Fig:app:CDF_drift_halt} for a cluster with  $\alpha_\nu = 10^{-2}$ and using the \citetalias{2021ApJ...910...51K} photoevaporation prescription.

Halting the dust evolution in these 5\% of discs naturally results in a cumulative distribution of the dust disc mass that extends to higher masses, as seen in Fig. \ref{Fig:app:CDF_drift_halt}. The thick lines shows the simulation including 5\% of discs with halted dust evolution, and the thin lines shows the standard model. This experiment illustrates how halting pebble drift in a fraction of the most massive discs can help explain the presence of very massive dust discs seen in evolved star forming regions, whilst the majority of discs agree well with being drift-dominated.

\end{appendix}

\
\end{document}